%% file: ms.tex
\documentclass[12pt]{article}
\pdfoutput=1
\usepackage[margin=1in]{geometry}
\usepackage{amsmath}
\usepackage{amssymb}
\usepackage[natbib=true,sorting=none]{biblatex}
\usepackage{tikz}
\usepackage{algorithm}
\usepackage{subcaption}
\usepackage{float}

\usepackage{pgfplots}
\usepackage{hyperref}

\pgfplotsset{compat=1.16}
\usetikzlibrary{backgrounds}
\usetikzlibrary{fit,positioning}
\usetikzlibrary{matrix}
\usetikzlibrary{shapes.geometric}
\usetikzlibrary{shapes.arrows}
\usetikzlibrary{arrows.meta}
\usetikzlibrary{intersections}
\usetikzlibrary{decorations.pathmorphing, shapes}

\usepackage{array}  
\DeclareMathOperator*{\argmax}{\text{argmax}}

\DeclareMathOperator*{\maximize}{\text{maximize}~}

\usepackage{enumitem} 
\usepackage{tikz} 
\usetikzlibrary{automata,arrows,shapes,decorations.pathreplacing} 
\usepackage{amssymb}
\usepackage{amsmath}
\usepackage{url}

\newcommand{\seqs}[1] {\mathcal{S}_{\uppercase{#1}}}
\newcommand{\subseq}[3] {#1_{#2:#3}}
\newcommand{\supp}[1] {\mathbf{supp}(#1)}
\newcommand{\R}{\mathbf{R}}
\newcommand{\concat}[2] {#1\cdot#2}
\newcommand{\vect}[1]{\R^{#1}}
\newcommand{\mat}[2]{\R^{#1\times #2 }}
\newcommand{\KE}{\mathbf{KE}}
\newcommand{\simplex}[1]{\Delta}

\title{Sequencing by Emergence: Modeling and Estimation}

\author{Nicholas Boyd, Samuel Woodhouse,  Kalim Mir}

\begin{document}
\maketitle

\begin{abstract}
Sequencing by Emergence (SEQE) is a new single-molecule nucleic acid (DNA/RNA) sequencing technology
that estimates sequence as an emergent property of the binding and localization of 
a repertoire of short oligonucleotide probes. 
SEQE promises to deliver accurate, ultra-long, haplotype-phased reads at the whole genome-scale for very low cost within 10 minutes.
The data SEQE generates requires entirely new inference techniques.
In this paper we introduce a probabilistic model of the 
SEQE measurement process and an algorithm that estimates sequence by solving a convex relaxation of the corresponding maximum likelihood problem.
We demonstrate the effectiveness of our algorithm on a variety of simulated datasets.
\end{abstract}

\section{Introduction}

Biological systems encode heritable information in the sequence of bases along the length of nucleic acids~\cite{watson1953molecular}. As a result,
reading nucleic acid sequences is crucial for countless scientific and medical processes. 
We propose an algorithm for \emph{Sequencing by Emergence} (SEQE)~\cite{mir2019sequencing, mir2020systems}, a new single-molecule sequencing technology potentially faster, more accurate, 
and less expensive than existing sequencing methods~\cite{bentley2008accurate, eid2009real, drmanac2010human, jain2016oxford}.
Notably, SEQE can generate megabase-long reads (and possibly whole chromosome lengths) at the same cost and time it takes to sequence just a few hundred bases; it is only constrained by the length of DNA molecules to which it is applied.
Additionally, SEQE has layers of built-in redundancy allowing the accuracy of reads to be increased arbitrarily by modestly increasing sequencing time.
The method is experimentally simple yet robust; it requires no library preparation, is completely enzyme-free and as such is low-cost and rapid.

The SEQE measurement process proceeds as follows. First, many molecules of DNA are elongated~\cite{parra1993high,bensimon1994alignment} (and typically aligned in one orientation) and attached to the surface of a microscope slide or cover glass.
Then, we record videos of the surface while different very short ($5$ nucleotide) fluorescently-tagged oligonucleotide (oligo) probes from a complete repertoire are flowed
over the sample DNA molecules.
In each video --- which we call a measurement --- multiple species of oligo probe are introduced, typically between 1 and 20 in number.
All probes are tagged with the exact same fluorescent label, so within a single round of imaging they are optically indistinguishable.
Because our probes are very short, they are extremely specific:
to a first approximation they hybridize only to their reverse complement in the DNA.
By this we mean they obey Watson-Crick base pairing: $\texttt{A}$ binds only to $\texttt{T}$ and $\texttt{C}$ binds only to $\texttt{G}$.
Unbound probes diffuse rapidly and thus appear as a uniform
background in the video; we use a total internal reflection microscope to reduce background intensity.
Under the correct experimental conditions probes will bind transiently:
hybridization sites along the DNA will appear to blink as probes bind and unbind repeatedly. 
As in the super-resolution imaging technique DNA-PAINT~\cite{jungmann2010single,schnitzbauer2017super}, 
if the number of probes bound in any single image is not too large, each hybridization event can be
localized individually and accurately (to within a few nanometers), 
despite the fact that the diffraction-induced point spread function of the microscope is much larger (typically 200-300nm).
A typical sequencing run consists of one to three hundred videos, each 5-60 seconds in length.

\begin{figure}
    \centering
    \begin{subfigure}[b]{0.4\textwidth}
        \centering
        \includegraphics[height=5cm]{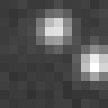}
        \caption{A single frame from a measurement.}
        \label{fig:frame}
    \end{subfigure}
    \hfill
    \begin{subfigure}[b]{0.4\textwidth}
        \centering
        \includegraphics[height=5cm]{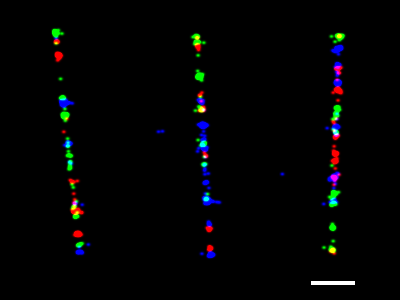}
        \caption{Localizations from three videos.}
        \label{fig:SR}
    \end{subfigure}
    \vfill
    \begin{subfigure}[b]{0.5\textwidth}
        \centering
        \includegraphics[width=\textwidth]{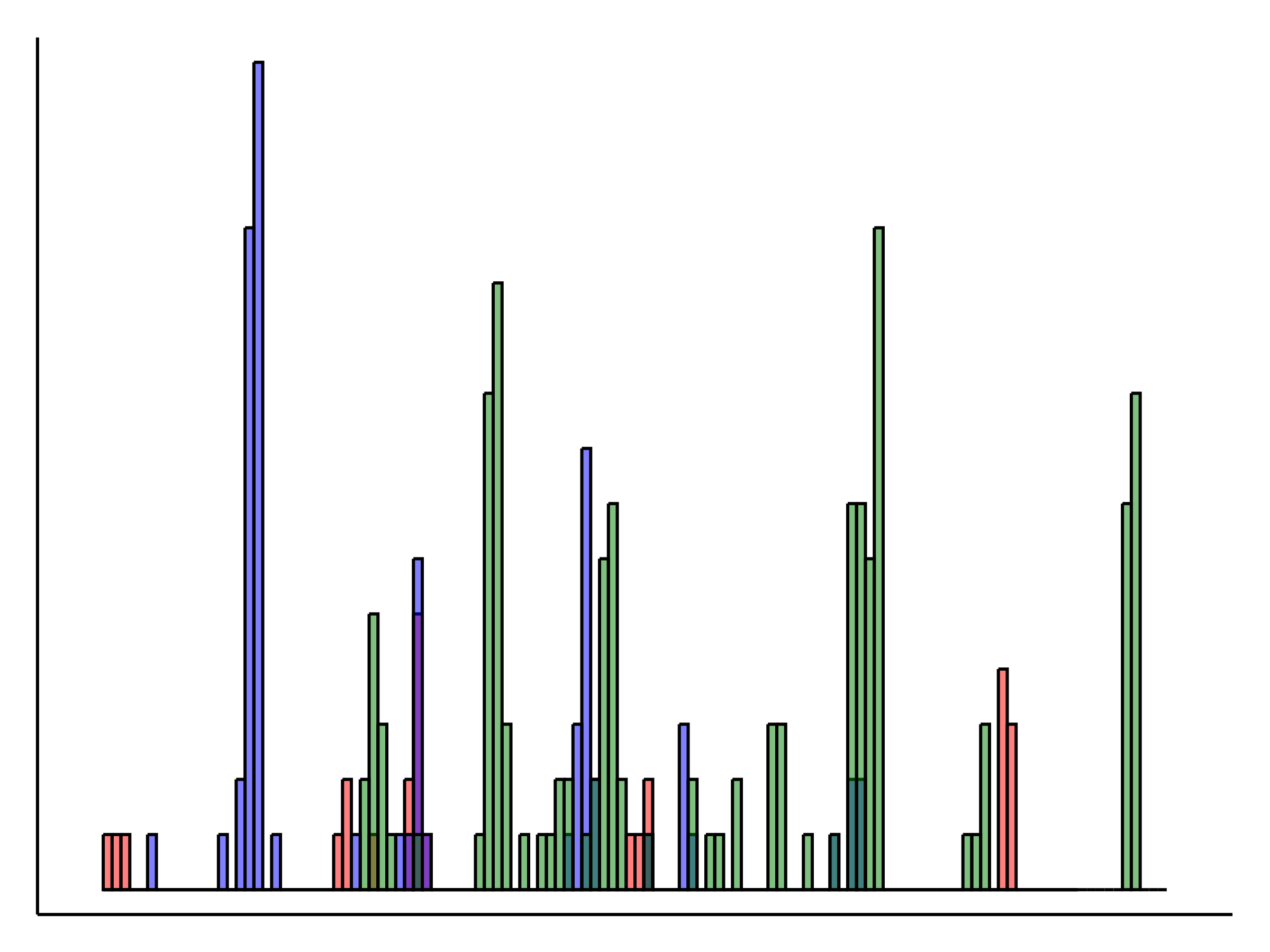}
        \caption{Histogram of 1D localizations for a single strand. }
        \label{fig:hist}
    \end{subfigure}
    
    \caption{A brief overview of the SEQE preprocessing steps. \ref{fig:frame} shows a single frame from a video before single molecule localization.
    \ref{fig:SR} shows the localizations from three measurements (each with a different set of probes) overlaid. The localizations are visualized by convolution with a small Gaussian kernel.
    Each round of imaging is displayed in a different color. 
    The scale bar is 100 nanometers, which corresponds to one pixel in \ref{fig:frame}.
    Three strands of DNA are visible. A histogram of localizations assigned and projected to the middle strand are displayed in~\ref{fig:hist}. Each bin is 5 nanometers wide.
    The input to the inference algorithms we describe in this paper are short sliding windows of 1D localizations similar to~\ref{fig:hist}.
    Note that while the field of view here and in our simulations is small (12 by 12 pixels), this is simply a matter of computational convenience.
    Modern TIRF microscopes can have 2048 by 2048 (or larger) pixel fields of view.}

       \label{fig:preprocessing}
\end{figure}

Before running our sequencing algorithm, several preprocessing steps are required.
First, each frame of each video is run through a single-molecule localization algorithm, generating a set of 2D coordinates (localizations) for each video.
Next, we identify all DNA strands in the sample and project localizations to the nearest strand. 
The final output of these pre-processing steps is a set of 1D localizations for each video for each segment of DNA. The algorithm we describe sequences each molecule independently, 
proceeding in a sliding-window fashion by processing regions corresponding to 150 nucleotides at a time.
See Figure~\ref{fig:preprocessing} for a graphical representation of these steps.

Even with the substantial resolution improvement afforded by single-molecule localization, the inverse problem is still quite difficult.
Each nucleotide of DNA is only 0.5 nanometers in length (after elongation); a hybridization event cannot be unambiguously mapped to a sequence position.
Additional complications arise because the probe species within a single video are indistinguishable, the probes are not perfectly specific, and the probes also occasionally bind 
to the microscope slide surface (even in the absence of DNA).

SEQE will be described in further detail in a forthcoming experimental paper, this paper focuses on modeling and inference for the data SEQE generates; 
as such we use simulated data based on realistic assumptions.
We propose a novel convex relaxation for a class of optimization problems over sequences that we successfully apply to the SEQE inverse problem. 

For clarity, for the remainder of this paper we assume all DNA is single-stranded; the models and algorithms we propose all extend easily to the double-stranded case. 

\paragraph{Outline}
\S\ref{s_problem} introduces the sequencing problem and our mathematical notation, while~\S\ref{PPP} and~\S\ref{s_relaxation} introduce a probabilistic 
model for the measurement process and an algorithm for the corresponding inverse problem, respectively. In~\S\ref{s_experiments} we 
illustrate the performance of SEQE and our algorithm in a series of simulated experiments. Finally, in~\S\ref{conclusion} we discuss limitations of this paper
along with potential improvements to the algorithms and models we present.


\section{Problem Statement and Notation}
\label{s_problem}
While each field of view contains many DNA molecules, each of which can be quite long, we process short (150nt) segments of each molecule independently --- it is easy to stitch the estimates for each segment together.
As such, the input data for the inference algorithm is a set of projected localizations for each of the $m$ measurements (a measurement is a single video)
--- each with the associated parameters of the experiment (probe mixture, concentrations, etc.).
That is, the data we concern ourselves with in the remainder of the paper are
\begin{equation*}
  D_1, \ldots, D_m \subset {\R}.
\end{equation*}
$x \in D_i$ means that we observed a localization in the $i$-th measurement $x$ nanometers from the beginning of the strand. 

The relationship between the sequence $s$ and the localizations $D_i$ is complex, but can be summarized as a combination of a few phenomena.
In each measurement, probes bind repeatedly to their reverse complements along the DNA strand. Localizations from each of these hybridization events will be displaced
from their actual physical location along the strand by some random distance due to errors in the single-molecule localization step --- typically on the order of a few nanometers.
There will be some level of
false positive localizations due to both non-specific hybridization to the DNA and transient interactions between the probes and the slide surface. 

The goal is to estimate the sequence $s$ from the experimental data $D_1, \ldots, D_m$.

\subsection{Mathematical Notation}
We denote the $i$-th standard basis vector, with length inferable from context, by $e_i$. The set of real numbers is $\R$.
The set of vectors of length $n$ with entries in $\R$ is $\vect{n}$; similarly, $\mat{m}{n}$ is the set of $m$ by $n$ real-valued matrices.

The set of length $N$ DNA sequences, $\{\texttt{A},\texttt{T},\texttt{C},\texttt{G}\}^N$, is $\seqs{n}$. For short sequences we use the letter $K$, and refer to such sequences as
$K$-mers (e.g. \texttt{ATCG} is a 4-mer). We denote the subsequence of a sequence $s$ from index $i$ to index $j$ (both inclusive) by $\subseq{s}{i}{j}$.
We write $\concat{s}{t}$ for the concatenation of $s$ and $t$.

Throughout, we interpret DNA sequences as integers: a sequence of length $N$ is associated with an integer between $1$ and $4^N$ by considering the sequence as a base-$4$ number.
As such, we may use sequences in contexts where integers are expected and vice versa.

\section{Probabilistic Model}
\label{PPP}

We approach sequencing by attempting to solve a maximum likelihood problem: 

\begin{equation}
\label{ML}
\maximize_{s \in \seqs{n}} \sum_{i=1}^m \log P_i(D_i | s).
\end{equation}

Here $\log P_i(D_i | s)$ (considered as a function of the sequence $s$) is the log-likelihood of the data $D_i$ under a probabilistic model we introduce.
We should note that we cannot solve~\eqref{ML} exactly; the combinatorial nature of the problem makes it quite difficult.
Instead, we introduce a heuristic algorithm to solve~\eqref{ML} approximately. In~\S\ref{s_experiments} we show that the heuristic algorithm works well, and often finds
sequences with higher likelihood than the ground truth sequence.

There is extensive work modeling and simulating the 
hybridization kinetics and optical measurement process in DNA-PAINT~\cite{jungmann2010single}. 
While these models are faithful to physical reality and easy to sample from --- in fact, we use them 
to generate our simulated data --- they do not
admit a tractable likelihood function $P_i(D_i | s)$. 
To get around this issue, we introduce an approximate probabilistic model that does have a tractable likelihood function:
we model the data generation process as a \emph{Poisson point process} (PPP). 

The PPP model is very simple. In short, for each measurement:

\begin{enumerate}
\item Each $K$-mer in the strand independently generates a random number of localizations.
\item The average number of localizations generated by a $K$-mer is a function of the measurement setup and the identity of the $K$-mer.
\item Each localization is randomly displaced from the true location of the $K$-mer.
\item A random number of uniformly distributed background localizations are added.
\item The final observation is the set of localizations falling within the sliding window $(l,u)$.
\end{enumerate}

\paragraph{Poisson point processes}

Before formally introducing the model, let us (briefly) review Poisson point processes.
A Poisson point process is a distribution on discrete subsets of $\R$ (collections of points) parameterized by a non-negative intensity measure $\lambda$.
The defining property of a PPP is that the number of points in a sample in a region $\mathcal{R} \subset \R$ is Poisson with mean $\lambda(\mathcal{R})$ and non-overlapping regions are independent.
There are two key properties of PPPs that we use. First, if the intensity measure $\lambda$ has a density (which we also call $\lambda$) with respect to Lebesgue measure,
the log-likelihood of $D \subset \R$ is a simple, concave functional of $\lambda$:

\begin{equation*} L(D| \lambda) = -\int \lambda(x) dx + \sum_{x \in D} \log \lambda(x). \end{equation*}
The second property we use is that the union of two independent samples from PPPs with intensity measures $\lambda_1$ and $\lambda_2$ is 
also distributed as a PPP with intensity measure $\lambda_1 + \lambda_2$.

\paragraph{Poisson point process model}
As a first step, we assume each site of length $K$ --- the length of the oligo probes, typically 5 --- along the DNA strand generates a number of hybridization events 
that is Poisson distributed and independent of every other site along the strand. Each of these events are detected with some fixed probability. 
This implies that the number of localizations generated by a $K$-mer $t$ is Poisson with mean given by a parameter of the experiment which we call $\theta^{(i)}(t)$ or the sensitivity
of experiment $i$:

\begin{equation*} \theta^{(i)} : \seqs{k} \rightarrow \mathbf{R} . \end{equation*}

We model the localization error induced by single-molecule localization as a displacement of each localization from its true position by a random 
distance sampled from an error distribution $e$. We take $e$ to be a Gaussian with $\sigma$ equal to the estimated 1D localization precision. Typically, $\sigma$ ranges from 1-10 nanometers.
This implies that the localizations generated by the single $K$-mer $t$ at location $l_j$ in the DNA strand are distributed according to a PPP with intensity
\begin{equation*} x \mapsto \theta^{(i)}(t) e(x | l_j), \end{equation*}
where $x$ is measured in nanometers and 
\begin{equation*} e(x|l_j) = \frac{1}{\sigma \sqrt{2\pi}}e^{-\left(\frac{x-l_j}{\sqrt{2}\sigma} \right)^2}.\end{equation*} 

The observed localizations are the union of localizations generated by all $K$-mers in the target sequence, which must then also be distributed according to a PPP,
with intensity
\[x \mapsto \sum_{j=1}^{N-K+1} \theta^{(i)}(\subseq{s}{j}{j+K-1}) e(x | l_j),\]
the sum of the intensities for each $K$-mer in the sequence.

We assume that localizations due to nonspecific binding to the slide surface are generated by a PPP with a uniform intensity per nanometer of $b$.
Additionally, because we run our algorithms on a sliding window between a lower cutoff $l$ and an upper limit $u$,
we multiply the intensity by the indicator function of the window $(l,u)$.

The final model (conditioned on the sequence $s$) for measurement $i$ is a PPP with intensity
\begin{equation*}
  \lambda_s^{(i)}(x) = \mathbf{1}_{(l,u)}(x)\left( b + \sum_{j=1}^{N-K+1} \theta^{(i)}(\subseq{s}{j}{j+K-1}) e(x | l_j) \right).
\end{equation*}

The total log-likelihood of the entire sequencing run is then:

\begin{equation*}
  \ell(s) = \sum_{i=1}^m -\int \lambda_s^{(i)}(x) dx + \sum_{x \in D_i} \log \lambda_s^{(i)}(x) .
\end{equation*}

The corresponding maximum-likelihood problem is
\begin{equation}
\label{denovo}
\maximize_{s \in \seqs{n}} \ell(s).
\end{equation}

To summarize, the parameters of the model are the sensitivities for each experiment, $\theta^{(1)}, \ldots, \theta^{(m)}$, the bounds $l$ and $u$, the background intensity $b$,
and the localization precision $\sigma$. The total number of parameters is roughly $m 4^K$.
The likelihood function depends on the parameters of the model and also the data, which are the observed localizations
$D_1, \ldots, D_m$.

Evaluation of $\ell$ is relatively cheap, $\mathcal{O}\left(N(\sum_{i=1}^m |D_i|) \right)$ floating point operations. This clearly varies with the experiment parameters, 
but as a rough guide, evaluating $\ell$ for an experiment with 200 rounds of imaging, each 10 seconds long, takes about 170 microseconds on a single thread of a laptop.

\paragraph{Inference}
The maximum likelihood problem is still difficult, however. Na\"{i}ve evaluation of $\ell$ for every possible sequence is clearly impossible:
$|\seqs{n}|$ is effectively infinite even for moderate values of $N$ (we use $N=150$).
Another option is a combinatorial algorithm that operates directly in the space of sequences.
For instance, we could try greedy search initialized to a random sequence. Greedy search checks the change of log-likelihood under all single-base modifications of the 
sequence (insertions, mutations, deletions) and
chooses the one that maximizes the increase; this is repeated until there are no edits that increase the likelihood. 
We show some results in~\S\ref{s_experiments} for this class of algorithms; in short, the greedy algorithm does not work on this problem --- 
though it's possible more sophisticated combinatorial algorithms could. 

Another option when faced with a difficult optimization problem is to solve a related but easier problem; we pursue this approach.

\section{Convex Relaxation}
\label{s_relaxation}

We relax~\eqref{denovo} into a convex problem, for which we propose an efficient algorithm.
We use the solution of this problem to generate approximate maximizers of the original objective.
For simplicity, in what follows we assume $\ell$ is the log-likelihood of a \emph{single} measurement; it should be a sum 
over measurements. Everything we discuss here can be trivially extended to such a sum.

First, consider searching for a probability distribution $p$ over $\seqs{n}$ instead of a single sequence.
It's impossible to represent general $p$ as $p \in \R^{4^N}$. However, if $p$ is sparse --- 
the support of $p$, $\supp{p} = \{s \in \seqs{n} : p_s > 0\}$, is small --- we can represent $p$ by storing only the nonzero entries
and values. Our algorithm only manipulates distributions on $\seqs{n}$ that have this sparse structure. 
Given a sparse $p$, we can 
quickly evaluate $\ell$ on all sequences in its support and take the sequence with maximal likelihood as an approximate solution to the maximum likelihood problem.

The question of how to generate a good distribution remains.
First, we note that an equivalent problem to~\eqref{denovo} is to find $p$ which maximizes $\ell$ in expectation:
\begin{equation}
  \label{not_relaxation}
  \maximize_{p \ge 0,~\sum_sp_s = 1} \sum_{s\in \seqs{n}}p_s \ell(\lambda_s).
\end{equation}
Here we overload $\ell$ as a function of the intensity measure $\lambda_s$.
A standard result holds that at least one solution to~\eqref{not_relaxation} is a point mass with objective value equal to the optimal value of~\eqref{denovo}:
 a distribution that puts all of its mass on $s_*$ where $s_*$ is a solution to the
original problem. While~\eqref{not_relaxation} is a convex problem, it is at least as hard to solve as~\eqref{denovo}.
To get a more tractable problem, we interchange the expectation and $\ell$:

\begin{equation}
  \label{relaxation}
  \maximize_{p \ge 0,~\sum_sp_s = 1} \ell\left(\sum_{s\in \seqs{n}}p_s \lambda_s \right).
\end{equation}
Because $\ell$, as a function of $\lambda_s$, is concave, this is a relaxation:
the optimal value of~\eqref{relaxation} upper-bounds the optimal value of the original optimization problem.
As such, a solution to~\eqref{relaxation} with exactly one nonzero coordinate, $e_s$, certifies that $s$ is optimal for the original discrete problem.
In our application, unfortunately, we rarely find such solutions. 
However, one can find solutions to the relaxed problem that are supported on very few sequences. Fortuitously, and perhaps surprisingly,
the support of these solutions often includes sequences which achieve very high objective values on the original, combinatorial problem. 

\section{The Conditional Gradient Method}
\label{s_cg}
While the relaxation proposed in~\S\ref{s_relaxation} is convex, it is quite difficult to solve because of the size of the decision variable $p$.
However, conditional gradient methods for such problems often find candidate solutions supported on very few coordinates and 
in practice materialize very few coordinates while running.

We briefly review the conditional gradient method (CGM) before describing its implementation on our problem.
The CGM solves the following convex problem:
\begin{equation*}
  \maximize_{x \in \mathcal{C}} f(x)
\end{equation*}
where $\mathcal{C}$ is a compact convex set and $f$ is a differentiable concave function. 
The CGM is an iterative algorithm that alternates between two steps: the conditional gradient step and the descent step.
In the conditional gradient step, we make a linear approximation to the function $f$ at the current iterate $x$ and maximize it over 
the constraint set:
\begin{equation*}
\maximize_{y \in \mathcal{C}} f(z) + \langle y-x, \nabla_z f(z) |_{z = x} \rangle.
\end{equation*}
The solution to this problem is a called a conditional gradient. 
Clearly, an efficient algorithm to solve this problem is required whenever one wishes to apply the CGM.

The descent step can take any of a number of forms, 
but the only requirement is that it increase the objective value sufficiently
(see~\cite{jaggi2013revisiting} for a detailed overview of conditional gradient methods).

In~\S\ref{ss_cg} and~\S\ref{ss_fc} we describe how we implement these steps for our problem.

\subsection{Conditional Gradients}
\label{ss_cg}

The crucial step in a conditional gradient method is, unsurprisingly, computing a conditional gradient,
which for~\eqref{relaxation} is any solution to the following optimization problem:
\begin{equation*}
\maximize_{y \ge 0,~\sum_sy_s = 1} \left\langle y, \nabla_p \ell\left.\left(\sum_{t\in \seqs{n}}p_t \lambda_t \right) \right\rvert_{p=p}\right\rangle
\end{equation*}
$p$ here is the current iterate. 

Because we're maximizing a linear functional over the probability simplex, there is at least one solution that is a point mass, so the problem simplifies to the following:

\begin{equation}
  \label{conditional_gradient_big}
\maximize_{s \in \seqs{n}} \left\langle e_s, \nabla_p \ell\left.\left(\sum_{t\in \seqs{n}}p_t \lambda_t \right) \right\rvert_{p=p}\right\rangle.
\end{equation}
We want to find the single sequence $s$ such that $e_s$ maximizes the directional derivative of $\ell$ at the current iterate, $p$.

As an aside, to gain some intuition about our approach one could imagine an extremely \emph{sharp} likelihood; in which case it seems reasonable that 
the linearized likelihood function in~\eqref{conditional_gradient_big} has a maximum at the ground truth sequence.
In this case the conditional gradient step actually solves the original discrete problem. 

Naïve solution of this problem is impossible: we can't even materialize the gradient.
However, due to structure in $\ell$, we can still solve~\eqref{conditional_gradient_big} efficiently --- in fact,
we show it is equivalent to a longest path problem in a small directed acyclic graph.

To do so we need to introduce the $K$-mer embedding of a sequence: $\KE : \seqs{n} \rightarrow \mat{4^K}{N-(K-1)}$.
$\KE$ maps a sequence $s$ to a sparse matrix with a single nonzero entry in each column: 
the concatenated one-hot embeddings of each $K$-mer in the sequence.
The columns of $\KE(s)$ are indexed by $K$-mers, while the rows are indexed by positions in the original sequence.
For particular position $j$ and $K$-mer $t$, $\KE(s)_{t,j}$ is $1$ if and only if $\subseq{s}{j}{j+(K-1)}$ is $t$.
In other words, we have
\begin{equation*}
\KE(s) = \sum_{j=1}^{N-(K-1)} e_{\subseq{s}{j}{j+K}}e_j^T.
\end{equation*}

Crucially, the intensity $\lambda_s$ is an \emph{affine} function of the $K$-mer embedding:
\begin{equation*}
  \lambda_s(x) = \mathbf{1}_{(l,u)}(x)\left( b + \sum_{j=1}^{N-K+1} \sum_{t=1}^{4^K} \KE(s)_{t,j} \theta(t) e(x | l_j) \right).
\end{equation*}
Note that this definition extends to all matrices in $\mat{4^K}{N-(K-1)}$ and that $\ell$ is concave as a function of the $K$-mer embedding.

If we overload $\ell$ to accept $\KE(s)$ (by composition with the affine operator described above) and 
introduce the linear operator $\mathcal{L} : \R^{4^N} \rightarrow \mat{4^K}{N-(K-1)}$ defined by
\begin{equation*} \mathcal{L}(p) = \sum_{s} p_s\KE(s) \end{equation*} we have
\begin{equation*}
  \ell\left(\sum_{s\in \seqs{n}}p_s \lambda_s \right) = \ell\left(\sum_{s \in \seqs{n}} p_s\KE(s)\right)  = \ell\left(\mathcal{L}(p)\right).
\end{equation*}

So, the conditional gradient problem is
\begin{equation*}
\maximize_{s \in \seqs{n}} \langle e_s, \nabla_p \ell(\mathcal{L}(p)) |_{p=p}\rangle.
\end{equation*}
Using the chain rule this is equivalent to
\begin{equation*}
\maximize_{s \in \seqs{n}} \langle \mathcal{L}(e_s), \nabla_z \ell(z) |_{z=\mathcal{L}(p)}\rangle.
\end{equation*}
Which in turn is equivalent to
\begin{equation}
  \label{cg}
\maximize_{s \in \seqs{n}} \langle \KE(s), G\rangle
\end{equation}
where $G = \nabla_z \ell(z) |_{z=\mathcal{L}(p)} \in \mat{4^K}{N-(K-1)}$.

Note that $G$ is easy to compute.
We need to find the sequence with $K$-mer embedding most closely correlated with the matrix $G$.
Luckily, there is an efficient dynamic program to find such a sequence.

Consider the directed acyclic graph, $B_{N,K}$, with vertices, $V$, given by the cartesian product of all $(K-1)$-mers, $\mathcal{S}_{K-1}$, and the set of positions $\{1, \ldots, N-(K-1)\}$.
The edge set, $E$, consists of all pairs $(s, i)$, $(t, j)$ such that $j = i+1$ and $s$ is valid prefix of $t$: $\subseq{s}{2}{K-1} = \subseq{t}{1}{K-2}$.
See Figure~\ref{fig:debruijn} for an example. 

\input{graph}

A sequence $s$ defines a unique path through this graph, starting at $(\subseq{s}{1}{K-1}, 1)$
and ending at $(\subseq{s}{N-K}{N}, N-(K-1))$ and traversing all the $(K-1)$-mers in the sequence.
Additionally, any path starting at a vertex in position 1 and ending at a vertex in position $N-(K-1)$ is associated with a unique sequence of length $N$. In other words,
such paths are in one-to-one correspondence with $\seqs{n}$. Again, see Figure~\ref{fig:debruijn} for an example.

Similarly, a matrix $G \in \mat{4^K}{N-(K-1)}$ is a packed representation of a set of weights on the edges of the graph:
the edge between $(s, i)$ and $(t,i+1)$ gets weight $G_{\concat{\subseq{s}{2}{K-1}}{\subseq{t}{K-1}{K-1}}, i}$.

The inner product in the objective of~\eqref{cg}, then, is simply the cost of the path defined by the sequence $s$ in $G_{N,K}$ with edge weights given by $D$.
This implies that~\eqref{cg} is a longest path problem in a directed acyclic graph which is efficiently computable by dynamic programming. 

In summary, despite the fact that naïve evaluation of the gradient is impossible, we can compute a conditional gradient efficiently in time $\mathcal{O}(N4^K)$.

\subsection{Descent Step}
\label{ss_fc}
We use the fully-corrective variant of the conditional gradient method; at each iteration we solve a restriction of the original optimization problem:
\begin{equation}
  \label{fc}
  \maximize_{p \ge 0,~\sum_sp_s = 1,~\supp{p} \subset \mathcal{I}} \ell(\mathcal{L}(p)).
\end{equation}
The set $\mathcal{I}$ is the support of the previous iterate with the addition of the single sequence returned by the conditional gradient step.
Note that if $\left| \mathcal{I} \right|$ is small, this problem is not too difficult: we simply re-parameterize so that the decision variable has dimension $\left| \mathcal{I} \right|$.
The motivation behind this choice is simple: if we initialize $p$ to $0$ --- technically not a feasible point, but this won't matter ---
the size of $\mathcal{I}$ is at most $k$, where $k$ is the number of iterations of the conditional gradient algorithm we're willing to run. We say at most $k$ because the solution
to~\eqref{fc} can, and in practice often does, have smaller support than $\mathcal{I}$.
In theory, there is even a global upper bound
on the size of $\mathcal{I}$, which for our problem is much smaller than $4^N$; though it's still much too large to matter in practice.
More importantly, in many practical applications $\mathcal{I}$ remains \emph{quite} small.

While standard algorithms can solve this problem, we use several tricks to speed up solution to~\eqref{fc}.
First, we warm-start using the previous value of $p$ to generate an initialization.
We use the proximal quasi-newton method with a BFGS approximation initialized to the true Hessian.
Finally, we solve the proximal subproblem using an active set method which attempts to guess the support of the solution and analytically solves the resulting equality-constrained quadratic program. If any of these methods fail we fall back to an interior point method.

\begin{algorithm}[H]
\caption{Conditional Gradient Method for~\eqref{relaxation}}
\label{alg_fc}
\smallskip
{\bf Repeat:}
\begin{enumerate}
\item Compute derivative: $G \leftarrow \nabla_z \ell(z) |_{z=\mathcal{L}(x)}.$
\item Compute conditional gradient: \[s \leftarrow \argmax_{s \in \seqs{n}} \langle \KE(s), G\rangle.\]
\item Solve fully-corrective subproblem: \[ x \leftarrow \argmax_{z \ge 0,~\sum_s z_s = 1,~\supp{z} \subset \supp{x} \cup \{s\}} \ell(\mathcal{L}(z)). \]
\end{enumerate}
\end{algorithm}

Conditional gradient methods generate a suboptimality bound at each iteration which we use to determine when to stop.

To turn a solution $p_*$ to~\eqref{relaxation} into potential solutions for~\eqref{denovo} we try two options. 
First, we compute $\ell(s)$ for each $s$ in the support of the solution. 
We also compute the objective value for the single sequence $s$ with $K$-mer embedding closest to $p_*$ in the euclidean metric 
(using the same algorithm we use to compute the conditional gradient).
We take as our point estimate the best of these sequences according to the objective function. 

\section{Simulations}
 \label{s_experiments}

In this section we present a few simulated experiments exploring the performance of SEQE and our inference algorithm in a variety of settings.
Code to reproduce these results will soon be made available.
All error bars are 95\% bootstrap confidence intervals computed using \href{https://github.com/juliangehring/Bootstrap.jl}{Bootstrap.jl}~\cite{julian_gehring}.

In all experiments we use an end-to-end, physically-realistic simulation of the measurement process; a continuous-time Markov model of the hybridization kinetics and a simple optics model of the microscope.
For single-molecule localization we use \href{https://github.com/nboyd/SingleMoleculeLocalization.jl}{SingleMoleculeLocalization.jl}.
The microscope simulation is the same in each experiment: each pixel is 100 nanometers in width (after magnification) with a 
12 by 12 pixel field of view and a 100ms exposure time.

To account for edge effects we simulate a single 450 nucleotide fragment of DNA and sequence the central 150 nucleotides.
We assume the stretching of the DNA is uniform with each nucleotide 0.5 nanometers in length. 
We set the window limits to 10 nanometers from the edges of this central 150nt segment.
Finally, we score our algorithms on how well the central 50 nucleotides of the reconstruction align to the ground truth --- alignments are computed using \href{https://github.com/BioJulia/BioAlignments.jl}{BioAlignments.jl}.
The per-base error metric we report is simply the number of mismatches, insertions, and deletions (divided by 50) in the alignment of the estimated sequence to the ground truth.

We use a very simple model of probe kinetics: probes bind to their reverse complement with a mean on-time of $0.3$ seconds and a mean off-time of $3$ seconds.
This is consistent with published DNA-PAINT experiments~\cite{civitci2020fast, schueder2019order}; it is 
generally possible to tune the experiment conditions to achieve similar kinetics for all probes.
We assume the on-time decays geometrically with the number of mismatches between the probe and the target at a rate of $1/20$ per mismatch. Likewise, the off-time increases geometrically at the same rate. 
We use a constant background localization distribution of $0.01$ localizations per nanometer.

For the PPP model, we use three parameters. We assume the sensitivity of a measurement is zero if the target $K$-mer is not in the mixture of probes used the in that measurement and a constant otherwise. 
The other two parameters are the localization precision $\sigma$ and the nonspecific binding rate $b$. Whenever we vary the simulation parameters (number of probes, signal-to-noise level, etc) we fit these three parameters using
coordinate ascent on the likelihood of additional simulated data. 

We run the CGM algorithm until the optimality gap (on the relaxed problem) is less than $0.01$.

In each simulation we choose a uniformly random collection of probe species for each measurement. 
As we discuss in~\S\ref{conclusion} this may result in a substantial \emph{overestimate} of the error rate compared to an optimized 
selection of oligos.

\paragraph{Greedy v.s. relaxation}

First, we compare the conditional gradient method to greedy combinatorial search.
Figure~\ref{fig_greedy} shows an example run of both algorithms on a simulated experiment, while Figure~\ref{fig_greedy_errors} shows
the error rates of the two algorithms on a series of simulated experiments. 
Clearly, the conditional gradient method applied to the relaxation substantially outperforms the greedy algorithm in terms of both computation time and final sequencing accuracy.

\begin{figure}[H]
  \centering
  \begin{subfigure}[b]{0.5\textwidth}
      \centering
      \input{"fig/fig_greedy.tex"}
\caption{Sample runs of the CGM and greedy algorithms on a simulated experiment. $\ell(s_*)$ is the likelihood of the ground truth sequence. For the CGM we display the likelihood of the best single sequence found; not the objective value on the relaxed problem.}
\label{fig_greedy}
  \end{subfigure}
  \hfill
  \begin{subfigure}[b]{0.5\textwidth}
      \centering
      \input{"fig/fig_greedy_errors.tex"}
\caption{Average per-nucleotide error rate within the aligned 50nt central segment for 100 runs of the CGM and greedy algorithms.
The (estimated) error rate of the CGM is zero.}
\label{fig_greedy_errors}
  \end{subfigure}
 
  
  \caption{Comparison between the greedy and CGM algorithms. We use 7 probes per measurement and 200 measurements each 15 seconds long.
  The localization precision is around 2.5 nanometers.}

     \label{fig:greedy}
\end{figure}

\paragraph{Multiplexing}
Next, in Figure~\ref{fig_rounds}, we investigate the impact of the multiplexing rate (the number of probes per measurement) on sequencing performance.
We deliberately choose a relatively low number of measurements (125) to highlight the impact of multiplexing.
If the number of probes is too low, sequencing fails because many parts of the target sequence are never hybridized.
If the number of probes is too high, localization errors increase as multiple probes are bound at the same time. 
Each localization also conveys less information about the sequence --- we cannot say which species of probe bound to produce the localization.
\begin{figure}[H]
\centering
\input{"fig/fig_probes.tex"}
\caption{Varying the number of probes per measurement in an experiment with 125 rounds of imaging, each 15 seconds long. 
Localization precision is about $2.5$nm. }
\label{fig_probes}
\end{figure}

\paragraph{Number of measurements}
Next, we hold the number of probes per measurement fixed at 7 and vary the number of measurements.
Each measurement is only 10 seconds long while the localization precision is about 2.8 nanometers. 
Figure~\ref{fig_rounds} shows that increasing the number of measurements dramatically increases sequencing performance.

\begin{figure}[H]
\centering
\input{"fig/fig_rounds.tex"}
\caption{Sequencing accuracy as a function of the number of measurements.
At 150 rounds we achieve an error rate less than $1\%$ in 25 minutes of imaging.}
\label{fig_rounds}
\end{figure}

\paragraph{Localization precision}

Finally, in Figure~\ref{fig_loc}, we vary the signal-to-noise ratio of the localizations while holding the 
other parameters fixed. In this experiment we use 7 probes per measurement and 150 measurements, each 15 seconds long.

\begin{figure}[H]
\centering
\input{"fig/fig_loc.tex"}
\caption{Two series of simulations demonstrate that localization precision strongly influences final sequencing accuracy.
Taking more or longer measurements can ameliorate loss of performance due to lower localization precision.}
\label{fig_loc}
\end{figure}

\section{Discussion and Future Work}
\label{conclusion}

\paragraph{Related sequencing technology}

SEQE is closely related to sequencing by hybridization (SbH)~\cite{drmanac1987yugoslav, southern1988united, lysov1988dna, pevzner19891, cantor1992report, southern1992analyzing, drmanac1993dna, drmanac1998accurate,
bains1988novel, bains1991hybridization, broude1994enhanced} which, as originally proposed, estimates a sequence from the $K$-mer spectrum: the set of $K$-mers that occur in the sequence.
An implementation of SbH by Sten Linnarsson and co-workers~\cite{pihlak2008rapid} gave a hint of the power of short, universal oligonucleotide probes.
SEQE shares with SBH the fact that multiple oligos hybridize to each base in the target sequence, adding redundancy. 
Standard SbH, however, is subject to extreme length limitations, indeed, the typically cited bound is that sequences can be no more than $2^K$
nucleotides long before the $K$-mer spectrum becomes insufficient to reconstruct the sequence ~\cite{dyer1994probability}.
The additional localization information available in SEQE means that the maximum sequence length is completely unbounded and unusual sequences
(i.e.\ repeats) present much less of an issue.

\paragraph{Limitations}

While the performance of SEQE on simulated data is quite impressive, there may be aspects that are harder in reality than as we have modeled.
We leave detailed discussion of SEQE to a forthcoming paper, but suffice it to say that
SEQE is quite flexible. If, for instance, it is difficult to achieve the relatively short off-times we use in our simulations
we could increase imaging time per measurement, 
increase the number of probes per measurement, or take more measurements. 
Additionally, the model we present can be easily modified to accommodate, for instance, variable stretching of the DNA.

On the computational and statistical side, we did not discuss many stages in the full sequencing pipeline.
For instance, strand identification and fitting, drift correction and alignment between measurements, and finally 
generating long reads by aligning the sliding windows. Additionally, all simulations in this paper are run on uniformly random DNA; real-world
sequences include difficult regions (e.g.\ homopolymer tracts) that may require specialized algorithms or measurements. 

While we did try several combinatorial algorithms in addition to the greedy algorithm presented in the simulations, it is possible that different combinatorial 
or purely heuristic algorithms could outperform the CGM in some measurement regimes.

\paragraph{Future work}

A clear next step is a better exploration of the large hyperparameter space of SEQE. For example, in 
the simulations in this paper the extent of probe multiplexing (the number of oligo species per imaging round) is limited by the short off-time 
of the probes. If we use probes with a longer dark time (by, for instance, lowering their concentration) this could allow for higher multiplexing rates.
Additionally, the probes we use in these simulations are very specific; but it's possible less specific probes could \emph{improve} sequencing
if we could exploit the different kinetics between perfect match hybridization and mismatched hybridization. 

One obvious improvement to SEQE that we did not discuss is multicolor imaging~\cite{wade2019124}.
By using multiple spectrally distinguishable dyes we could run multiple measurements concurrently. 
Existing work in DNA-PAINT has demonstrated at least 3 color imaging; for SEQE this translates to a 3x decrease in sequencing time.
This could mean a total imaging time of less than 10 minutes. Several additional strategies exist to further decrease the imaging time. 

Another improvement would be the intelligent selection of the oligo species deployed in multiplex within each measurement.
In the simulations presented in this paper, probe species are chosen randomly,
but it is likely that sequencing performance can be improved by, for instance, choosing species that are unlikely to overlap on the target.

On the inference side, there are several avenues for improvement. 
First, our current model does not use kinetic information other than the total number of localizations generated by a particular probe, while we 
could potentially use the dwell times of various probes directly in inference.
This could be easily accomplished by using a PPP with support on the cartesian product of the spatial position and the measured dwell time.
Additional improvements could be possible by combining various stages of the inference pipeline:
for instance, we could use the fact that localizations occur along strands during localization, or build an inference model that can handle raw image data.

Further improvements to the algorithm are possible. For instance, in many applications it may be desirable to compute an approximate
posterior distribution instead of a point estimate. In another direction, the analysis in~\S\ref{ss_cg} shows that 
it may be possible to optimize directly over the convex hull of all possible $K$-mer embeddings;
indeed, this set is the unit ball of an atomic norm and the flow-polytope of the graph $B_{N,K}$.

Finally, the SEQE process, PPP model, and the algorithm we propose here can also be applied to sequencing with a reference as a prior.
This problem is substantially easier and requires less total imaging time. 

\printbibliography
\end{document}

%% file: graph
\begin{figure}
\begin{center}
  \resizebox{1.0\textwidth}{!}{
\begin{tikzpicture}[xscale=1.5,yscale=7.5]
\node[circle,fill=none,draw=black,minimum size=4mm,ultra thick] (AA_1) at (1,1) {\scriptsize(\texttt{AA}, 1)};
\node[circle,fill=none,draw=black,minimum size=4mm] (AA_2) at (1,2) {\scriptsize(\texttt{AA}, 2)};
\node[circle,fill=none,draw=black,minimum size=4mm] (AA_3) at (1,3) {\scriptsize(\texttt{AA}, 3)};
\node[circle,fill=none,draw=black,minimum size=4mm] (AA_4) at (1,4) {\scriptsize(\texttt{AA}, 4)};
\node[circle,fill=none,draw=black,minimum size=4mm] (AA_5) at (1,5) {\scriptsize(\texttt{AA}, 5)};
\node[circle,fill=none,draw=black,minimum size=4mm] (AC_1) at (2,1) {\scriptsize(\texttt{AC}, 1)};
\node[circle,fill=none,draw=black,minimum size=4mm] (AC_2) at (2,2) {\scriptsize(\texttt{AC}, 2)};
\node[circle,fill=none,draw=black,minimum size=4mm] (AC_3) at (2,3) {\scriptsize(\texttt{AC}, 3)};
\node[circle,fill=none,draw=black,minimum size=4mm] (AC_4) at (2,4) {\scriptsize(\texttt{AC}, 4)};
\node[circle,fill=none,draw=black,minimum size=4mm] (AC_5) at (2,5) {\scriptsize(\texttt{AC}, 5)};
\node[circle,fill=none,draw=black,minimum size=4mm] (AG_1) at (3,1) {\scriptsize(\texttt{AG}, 1)};
\node[circle,fill=none,draw=black,minimum size=4mm] (AG_2) at (3,2) {\scriptsize(\texttt{AG}, 2)};
\node[circle,fill=none,draw=black,minimum size=4mm] (AG_3) at (3,3) {\scriptsize(\texttt{AG}, 3)};
\node[circle,fill=none,draw=black,minimum size=4mm,ultra thick] (AG_4) at (3,4) {\scriptsize(\texttt{AG}, 4)};
\node[circle,fill=none,draw=black,minimum size=4mm] (AG_5) at (3,5) {\scriptsize(\texttt{AG}, 5)};
\node[circle,fill=none,draw=black,minimum size=4mm] (AT_1) at (4,1) {\scriptsize(\texttt{AT}, 1)};
\node[circle,fill=none,draw=black,minimum size=4mm,ultra thick] (AT_2) at (4,2) {\scriptsize(\texttt{AT}, 2)};
\node[circle,fill=none,draw=black,minimum size=4mm] (AT_3) at (4,3) {\scriptsize(\texttt{AT}, 3)};
\node[circle,fill=none,draw=black,minimum size=4mm] (AT_4) at (4,4) {\scriptsize(\texttt{AT}, 4)};
\node[circle,fill=none,draw=black,minimum size=4mm] (AT_5) at (4,5) {\scriptsize(\texttt{AT}, 5)};
\node[circle,fill=none,draw=black,minimum size=4mm] (CA_1) at (5,1) {\scriptsize(\texttt{CA}, 1)};
\node[circle,fill=none,draw=black,minimum size=4mm] (CA_2) at (5,2) {\scriptsize(\texttt{CA}, 2)};
\node[circle,fill=none,draw=black,minimum size=4mm] (CA_3) at (5,3) {\scriptsize(\texttt{CA}, 3)};
\node[circle,fill=none,draw=black,minimum size=4mm] (CA_4) at (5,4) {\scriptsize(\texttt{CA}, 4)};
\node[circle,fill=none,draw=black,minimum size=4mm] (CA_5) at (5,5) {\scriptsize(\texttt{CA}, 5)};
\node[circle,fill=none,draw=black,minimum size=4mm] (CC_1) at (6,1) {\scriptsize(\texttt{CC}, 1)};
\node[circle,fill=none,draw=black,minimum size=4mm] (CC_2) at (6,2) {\scriptsize(\texttt{CC}, 2)};
\node[circle,fill=none,draw=black,minimum size=4mm] (CC_3) at (6,3) {\scriptsize(\texttt{CC}, 3)};
\node[circle,fill=none,draw=black,minimum size=4mm] (CC_4) at (6,4) {\scriptsize(\texttt{CC}, 4)};
\node[circle,fill=none,draw=black,minimum size=4mm] (CC_5) at (6,5) {\scriptsize(\texttt{CC}, 5)};
\node[circle,fill=none,draw=black,minimum size=4mm] (CG_1) at (7,1) {\scriptsize(\texttt{CG}, 1)};
\node[circle,fill=none,draw=black,minimum size=4mm] (CG_2) at (7,2) {\scriptsize(\texttt{CG}, 2)};
\node[circle,fill=none,draw=black,minimum size=4mm] (CG_3) at (7,3) {\scriptsize(\texttt{CG}, 3)};
\node[circle,fill=none,draw=black,minimum size=4mm] (CG_4) at (7,4) {\scriptsize(\texttt{CG}, 4)};
\node[circle,fill=none,draw=black,minimum size=4mm] (CG_5) at (7,5) {\scriptsize(\texttt{CG}, 5)};
\node[circle,fill=none,draw=black,minimum size=4mm] (CT_1) at (8,1) {\scriptsize(\texttt{CT}, 1)};
\node[circle,fill=none,draw=black,minimum size=4mm] (CT_2) at (8,2) {\scriptsize(\texttt{CT}, 2)};
\node[circle,fill=none,draw=black,minimum size=4mm] (CT_3) at (8,3) {\scriptsize(\texttt{CT}, 3)};
\node[circle,fill=none,draw=black,minimum size=4mm] (CT_4) at (8,4) {\scriptsize(\texttt{CT}, 4)};
\node[circle,fill=none,draw=black,minimum size=4mm] (CT_5) at (8,5) {\scriptsize(\texttt{CT}, 5)};
\node[circle,fill=none,draw=black,minimum size=4mm] (GA_1) at (9,1) {\scriptsize(\texttt{GA}, 1)};
\node[circle,fill=none,draw=black,minimum size=4mm] (GA_2) at (9,2) {\scriptsize(\texttt{GA}, 2)};
\node[circle,fill=none,draw=black,minimum size=4mm] (GA_3) at (9,3) {\scriptsize(\texttt{GA}, 3)};
\node[circle,fill=none,draw=black,minimum size=4mm] (GA_4) at (9,4) {\scriptsize(\texttt{GA}, 4)};
\node[circle,fill=none,draw=black,minimum size=4mm,ultra thick] (GA_5) at (9,5) {\scriptsize(\texttt{GA}, 5)};
\node[circle,fill=none,draw=black,minimum size=4mm] (GC_1) at (10,1) {\scriptsize(\texttt{GC}, 1)};
\node[circle,fill=none,draw=black,minimum size=4mm] (GC_2) at (10,2) {\scriptsize(\texttt{GC}, 2)};
\node[circle,fill=none,draw=black,minimum size=4mm] (GC_3) at (10,3) {\scriptsize(\texttt{GC}, 3)};
\node[circle,fill=none,draw=black,minimum size=4mm] (GC_4) at (10,4) {\scriptsize(\texttt{GC}, 4)};
\node[circle,fill=none,draw=black,minimum size=4mm] (GC_5) at (10,5) {\scriptsize(\texttt{GC}, 5)};
\node[circle,fill=none,draw=black,minimum size=4mm] (GG_1) at (11,1) {\scriptsize(\texttt{GG}, 1)};
\node[circle,fill=none,draw=black,minimum size=4mm] (GG_2) at (11,2) {\scriptsize(\texttt{GG}, 2)};
\node[circle,fill=none,draw=black,minimum size=4mm] (GG_3) at (11,3) {\scriptsize(\texttt{GG}, 3)};
\node[circle,fill=none,draw=black,minimum size=4mm] (GG_4) at (11,4) {\scriptsize(\texttt{GG}, 4)};
\node[circle,fill=none,draw=black,minimum size=4mm] (GG_5) at (11,5) {\scriptsize(\texttt{GG}, 5)};
\node[circle,fill=none,draw=black,minimum size=4mm] (GT_1) at (12,1) {\scriptsize(\texttt{GT}, 1)};
\node[circle,fill=none,draw=black,minimum size=4mm] (GT_2) at (12,2) {\scriptsize(\texttt{GT}, 2)};
\node[circle,fill=none,draw=black,minimum size=4mm] (GT_3) at (12,3) {\scriptsize(\texttt{GT}, 3)};
\node[circle,fill=none,draw=black,minimum size=4mm] (GT_4) at (12,4) {\scriptsize(\texttt{GT}, 4)};
\node[circle,fill=none,draw=black,minimum size=4mm] (GT_5) at (12,5) {\scriptsize(\texttt{GT}, 5)};
\node[circle,fill=none,draw=black,minimum size=4mm] (TA_1) at (13,1) {\scriptsize(\texttt{TA}, 1)};
\node[circle,fill=none,draw=black,minimum size=4mm] (TA_2) at (13,2) {\scriptsize(\texttt{TA}, 2)};
\node[circle,fill=none,draw=black,minimum size=4mm,ultra thick] (TA_3) at (13,3) {\scriptsize(\texttt{TA}, 3)};
\node[circle,fill=none,draw=black,minimum size=4mm] (TA_4) at (13,4) {\scriptsize(\texttt{TA}, 4)};
\node[circle,fill=none,draw=black,minimum size=4mm] (TA_5) at (13,5) {\scriptsize(\texttt{TA}, 5)};
\node[circle,fill=none,draw=black,minimum size=4mm] (TC_1) at (14,1) {\scriptsize(\texttt{TC}, 1)};
\node[circle,fill=none,draw=black,minimum size=4mm] (TC_2) at (14,2) {\scriptsize(\texttt{TC}, 2)};
\node[circle,fill=none,draw=black,minimum size=4mm] (TC_3) at (14,3) {\scriptsize(\texttt{TC}, 3)};
\node[circle,fill=none,draw=black,minimum size=4mm] (TC_4) at (14,4) {\scriptsize(\texttt{TC}, 4)};
\node[circle,fill=none,draw=black,minimum size=4mm] (TC_5) at (14,5) {\scriptsize(\texttt{TC}, 5)};
\node[circle,fill=none,draw=black,minimum size=4mm] (TG_1) at (15,1) {\scriptsize(\texttt{TG}, 1)};
\node[circle,fill=none,draw=black,minimum size=4mm] (TG_2) at (15,2) {\scriptsize(\texttt{TG}, 2)};
\node[circle,fill=none,draw=black,minimum size=4mm] (TG_3) at (15,3) {\scriptsize(\texttt{TG}, 3)};
\node[circle,fill=none,draw=black,minimum size=4mm] (TG_4) at (15,4) {\scriptsize(\texttt{TG}, 4)};
\node[circle,fill=none,draw=black,minimum size=4mm] (TG_5) at (15,5) {\scriptsize(\texttt{TG}, 5)};
\node[circle,fill=none,draw=black,minimum size=4mm] (TT_1) at (16,1) {\scriptsize(\texttt{TT}, 1)};
\node[circle,fill=none,draw=black,minimum size=4mm] (TT_2) at (16,2) {\scriptsize(\texttt{TT}, 2)};
\node[circle,fill=none,draw=black,minimum size=4mm] (TT_3) at (16,3) {\scriptsize(\texttt{TT}, 3)};
\node[circle,fill=none,draw=black,minimum size=4mm] (TT_4) at (16,4) {\scriptsize(\texttt{TT}, 4)};
\node[circle,fill=none,draw=black,minimum size=4mm] (TT_5) at (16,5) {\scriptsize(\texttt{TT}, 5)};
\draw[-{Stealth[scale=1.4,angle'=45]}] (AA_1) to (AA_2);
\draw[-{Stealth[scale=1.4,angle'=45]}] (AA_1) to (AC_2);
\draw[-{Stealth[scale=1.4,angle'=45]}] (AA_1) to (AG_2);
\draw[-{Stealth[scale=1.4,angle'=45]}, ultra thick] (AA_1) to (AT_2);
\draw[-{Stealth[scale=1.4,angle'=45]}] (AC_1) to (CA_2);
\draw[-{Stealth[scale=1.4,angle'=45]}] (AC_1) to (CC_2);
\draw[-{Stealth[scale=1.4,angle'=45]}] (AC_1) to (CG_2);
\draw[-{Stealth[scale=1.4,angle'=45]}] (AC_1) to (CT_2);
\draw[-{Stealth[scale=1.4,angle'=45]}] (AG_1) to (GA_2);
\draw[-{Stealth[scale=1.4,angle'=45]}] (AG_1) to (GC_2);
\draw[-{Stealth[scale=1.4,angle'=45]}] (AG_1) to (GG_2);
\draw[-{Stealth[scale=1.4,angle'=45]}] (AG_1) to (GT_2);
\draw[-{Stealth[scale=1.4,angle'=45]}] (AT_1) to (TA_2);
\draw[-{Stealth[scale=1.4,angle'=45]}] (AT_1) to (TC_2);
\draw[-{Stealth[scale=1.4,angle'=45]}] (AT_1) to (TG_2);
\draw[-{Stealth[scale=1.4,angle'=45]}] (AT_1) to (TT_2);
\draw[-{Stealth[scale=1.4,angle'=45]}] (CA_1) to (AA_2);
\draw[-{Stealth[scale=1.4,angle'=45]}] (CA_1) to (AC_2);
\draw[-{Stealth[scale=1.4,angle'=45]}] (CA_1) to (AG_2);
\draw[-{Stealth[scale=1.4,angle'=45]}] (CA_1) to (AT_2);
\draw[-{Stealth[scale=1.4,angle'=45]}] (CC_1) to (CA_2);
\draw[-{Stealth[scale=1.4,angle'=45]}] (CC_1) to (CC_2);
\draw[-{Stealth[scale=1.4,angle'=45]}] (CC_1) to (CG_2);
\draw[-{Stealth[scale=1.4,angle'=45]}] (CC_1) to (CT_2);
\draw[-{Stealth[scale=1.4,angle'=45]}] (CG_1) to (GA_2);
\draw[-{Stealth[scale=1.4,angle'=45]}] (CG_1) to (GC_2);
\draw[-{Stealth[scale=1.4,angle'=45]}] (CG_1) to (GG_2);
\draw[-{Stealth[scale=1.4,angle'=45]}] (CG_1) to (GT_2);
\draw[-{Stealth[scale=1.4,angle'=45]}] (CT_1) to (TA_2);
\draw[-{Stealth[scale=1.4,angle'=45]}] (CT_1) to (TC_2);
\draw[-{Stealth[scale=1.4,angle'=45]}] (CT_1) to (TG_2);
\draw[-{Stealth[scale=1.4,angle'=45]}] (CT_1) to (TT_2);
\draw[-{Stealth[scale=1.4,angle'=45]}] (GA_1) to (AA_2);
\draw[-{Stealth[scale=1.4,angle'=45]}] (GA_1) to (AC_2);
\draw[-{Stealth[scale=1.4,angle'=45]}] (GA_1) to (AG_2);
\draw[-{Stealth[scale=1.4,angle'=45]}] (GA_1) to (AT_2);
\draw[-{Stealth[scale=1.4,angle'=45]}] (GC_1) to (CA_2);
\draw[-{Stealth[scale=1.4,angle'=45]}] (GC_1) to (CC_2);
\draw[-{Stealth[scale=1.4,angle'=45]}] (GC_1) to (CG_2);
\draw[-{Stealth[scale=1.4,angle'=45]}] (GC_1) to (CT_2);
\draw[-{Stealth[scale=1.4,angle'=45]}] (GG_1) to (GA_2);
\draw[-{Stealth[scale=1.4,angle'=45]}] (GG_1) to (GC_2);
\draw[-{Stealth[scale=1.4,angle'=45]}] (GG_1) to (GG_2);
\draw[-{Stealth[scale=1.4,angle'=45]}] (GG_1) to (GT_2);
\draw[-{Stealth[scale=1.4,angle'=45]}] (GT_1) to (TA_2);
\draw[-{Stealth[scale=1.4,angle'=45]}] (GT_1) to (TC_2);
\draw[-{Stealth[scale=1.4,angle'=45]}] (GT_1) to (TG_2);
\draw[-{Stealth[scale=1.4,angle'=45]}] (GT_1) to (TT_2);
\draw[-{Stealth[scale=1.4,angle'=45]}] (TA_1) to (AA_2);
\draw[-{Stealth[scale=1.4,angle'=45]}] (TA_1) to (AC_2);
\draw[-{Stealth[scale=1.4,angle'=45]}] (TA_1) to (AG_2);
\draw[-{Stealth[scale=1.4,angle'=45]}] (TA_1) to (AT_2);
\draw[-{Stealth[scale=1.4,angle'=45]}] (TC_1) to (CA_2);
\draw[-{Stealth[scale=1.4,angle'=45]}] (TC_1) to (CC_2);
\draw[-{Stealth[scale=1.4,angle'=45]}] (TC_1) to (CG_2);
\draw[-{Stealth[scale=1.4,angle'=45]}] (TC_1) to (CT_2);
\draw[-{Stealth[scale=1.4,angle'=45]}] (TG_1) to (GA_2);
\draw[-{Stealth[scale=1.4,angle'=45]}] (TG_1) to (GC_2);
\draw[-{Stealth[scale=1.4,angle'=45]}] (TG_1) to (GG_2);
\draw[-{Stealth[scale=1.4,angle'=45]}] (TG_1) to (GT_2);
\draw[-{Stealth[scale=1.4,angle'=45]}] (TT_1) to (TA_2);
\draw[-{Stealth[scale=1.4,angle'=45]}] (TT_1) to (TC_2);
\draw[-{Stealth[scale=1.4,angle'=45]}] (TT_1) to (TG_2);
\draw[-{Stealth[scale=1.4,angle'=45]}] (TT_1) to (TT_2);
\draw[-{Stealth[scale=1.4,angle'=45]}] (AA_2) to (AA_3);
\draw[-{Stealth[scale=1.4,angle'=45]}] (AA_2) to (AC_3);
\draw[-{Stealth[scale=1.4,angle'=45]}] (AA_2) to (AG_3);
\draw[-{Stealth[scale=1.4,angle'=45]}] (AA_2) to (AT_3);
\draw[-{Stealth[scale=1.4,angle'=45]}] (AC_2) to (CA_3);
\draw[-{Stealth[scale=1.4,angle'=45]}] (AC_2) to (CC_3);
\draw[-{Stealth[scale=1.4,angle'=45]}] (AC_2) to (CG_3);
\draw[-{Stealth[scale=1.4,angle'=45]}] (AC_2) to (CT_3);
\draw[-{Stealth[scale=1.4,angle'=45]}] (AG_2) to (GA_3);
\draw[-{Stealth[scale=1.4,angle'=45]}] (AG_2) to (GC_3);
\draw[-{Stealth[scale=1.4,angle'=45]}] (AG_2) to (GG_3);
\draw[-{Stealth[scale=1.4,angle'=45]}] (AG_2) to (GT_3);
\draw[-{Stealth[scale=1.4,angle'=45]}, ultra thick] (AT_2) to (TA_3);
\draw[-{Stealth[scale=1.4,angle'=45]}] (AT_2) to (TC_3);
\draw[-{Stealth[scale=1.4,angle'=45]}] (AT_2) to (TG_3);
\draw[-{Stealth[scale=1.4,angle'=45]}] (AT_2) to (TT_3);
\draw[-{Stealth[scale=1.4,angle'=45]}] (CA_2) to (AA_3);
\draw[-{Stealth[scale=1.4,angle'=45]}] (CA_2) to (AC_3);
\draw[-{Stealth[scale=1.4,angle'=45]}] (CA_2) to (AG_3);
\draw[-{Stealth[scale=1.4,angle'=45]}] (CA_2) to (AT_3);
\draw[-{Stealth[scale=1.4,angle'=45]}] (CC_2) to (CA_3);
\draw[-{Stealth[scale=1.4,angle'=45]}] (CC_2) to (CC_3);
\draw[-{Stealth[scale=1.4,angle'=45]}] (CC_2) to (CG_3);
\draw[-{Stealth[scale=1.4,angle'=45]}] (CC_2) to (CT_3);
\draw[-{Stealth[scale=1.4,angle'=45]}] (CG_2) to (GA_3);
\draw[-{Stealth[scale=1.4,angle'=45]}] (CG_2) to (GC_3);
\draw[-{Stealth[scale=1.4,angle'=45]}] (CG_2) to (GG_3);
\draw[-{Stealth[scale=1.4,angle'=45]}] (CG_2) to (GT_3);
\draw[-{Stealth[scale=1.4,angle'=45]}] (CT_2) to (TA_3);
\draw[-{Stealth[scale=1.4,angle'=45]}] (CT_2) to (TC_3);
\draw[-{Stealth[scale=1.4,angle'=45]}] (CT_2) to (TG_3);
\draw[-{Stealth[scale=1.4,angle'=45]}] (CT_2) to (TT_3);
\draw[-{Stealth[scale=1.4,angle'=45]}] (GA_2) to (AA_3);
\draw[-{Stealth[scale=1.4,angle'=45]}] (GA_2) to (AC_3);
\draw[-{Stealth[scale=1.4,angle'=45]}] (GA_2) to (AG_3);
\draw[-{Stealth[scale=1.4,angle'=45]}] (GA_2) to (AT_3);
\draw[-{Stealth[scale=1.4,angle'=45]}] (GC_2) to (CA_3);
\draw[-{Stealth[scale=1.4,angle'=45]}] (GC_2) to (CC_3);
\draw[-{Stealth[scale=1.4,angle'=45]}] (GC_2) to (CG_3);
\draw[-{Stealth[scale=1.4,angle'=45]}] (GC_2) to (CT_3);
\draw[-{Stealth[scale=1.4,angle'=45]}] (GG_2) to (GA_3);
\draw[-{Stealth[scale=1.4,angle'=45]}] (GG_2) to (GC_3);
\draw[-{Stealth[scale=1.4,angle'=45]}] (GG_2) to (GG_3);
\draw[-{Stealth[scale=1.4,angle'=45]}] (GG_2) to (GT_3);
\draw[-{Stealth[scale=1.4,angle'=45]}] (GT_2) to (TA_3);
\draw[-{Stealth[scale=1.4,angle'=45]}] (GT_2) to (TC_3);
\draw[-{Stealth[scale=1.4,angle'=45]}] (GT_2) to (TG_3);
\draw[-{Stealth[scale=1.4,angle'=45]}] (GT_2) to (TT_3);
\draw[-{Stealth[scale=1.4,angle'=45]}] (TA_2) to (AA_3);
\draw[-{Stealth[scale=1.4,angle'=45]}] (TA_2) to (AC_3);
\draw[-{Stealth[scale=1.4,angle'=45]}] (TA_2) to (AG_3);
\draw[-{Stealth[scale=1.4,angle'=45]}] (TA_2) to (AT_3);
\draw[-{Stealth[scale=1.4,angle'=45]}] (TC_2) to (CA_3);
\draw[-{Stealth[scale=1.4,angle'=45]}] (TC_2) to (CC_3);
\draw[-{Stealth[scale=1.4,angle'=45]}] (TC_2) to (CG_3);
\draw[-{Stealth[scale=1.4,angle'=45]}] (TC_2) to (CT_3);
\draw[-{Stealth[scale=1.4,angle'=45]}] (TG_2) to (GA_3);
\draw[-{Stealth[scale=1.4,angle'=45]}] (TG_2) to (GC_3);
\draw[-{Stealth[scale=1.4,angle'=45]}] (TG_2) to (GG_3);
\draw[-{Stealth[scale=1.4,angle'=45]}] (TG_2) to (GT_3);
\draw[-{Stealth[scale=1.4,angle'=45]}] (TT_2) to (TA_3);
\draw[-{Stealth[scale=1.4,angle'=45]}] (TT_2) to (TC_3);
\draw[-{Stealth[scale=1.4,angle'=45]}] (TT_2) to (TG_3);
\draw[-{Stealth[scale=1.4,angle'=45]}] (TT_2) to (TT_3);
\draw[-{Stealth[scale=1.4,angle'=45]}] (AA_3) to (AA_4);
\draw[-{Stealth[scale=1.4,angle'=45]}] (AA_3) to (AC_4);
\draw[-{Stealth[scale=1.4,angle'=45]}] (AA_3) to (AG_4);
\draw[-{Stealth[scale=1.4,angle'=45]}] (AA_3) to (AT_4);
\draw[-{Stealth[scale=1.4,angle'=45]}] (AC_3) to (CA_4);
\draw[-{Stealth[scale=1.4,angle'=45]}] (AC_3) to (CC_4);
\draw[-{Stealth[scale=1.4,angle'=45]}] (AC_3) to (CG_4);
\draw[-{Stealth[scale=1.4,angle'=45]}] (AC_3) to (CT_4);
\draw[-{Stealth[scale=1.4,angle'=45]}] (AG_3) to (GA_4);
\draw[-{Stealth[scale=1.4,angle'=45]}] (AG_3) to (GC_4);
\draw[-{Stealth[scale=1.4,angle'=45]}] (AG_3) to (GG_4);
\draw[-{Stealth[scale=1.4,angle'=45]}] (AG_3) to (GT_4);
\draw[-{Stealth[scale=1.4,angle'=45]}] (AT_3) to (TA_4);
\draw[-{Stealth[scale=1.4,angle'=45]}] (AT_3) to (TC_4);
\draw[-{Stealth[scale=1.4,angle'=45]}] (AT_3) to (TG_4);
\draw[-{Stealth[scale=1.4,angle'=45]}] (AT_3) to (TT_4);
\draw[-{Stealth[scale=1.4,angle'=45]}] (CA_3) to (AA_4);
\draw[-{Stealth[scale=1.4,angle'=45]}] (CA_3) to (AC_4);
\draw[-{Stealth[scale=1.4,angle'=45]}] (CA_3) to (AG_4);
\draw[-{Stealth[scale=1.4,angle'=45]}] (CA_3) to (AT_4);
\draw[-{Stealth[scale=1.4,angle'=45]}] (CC_3) to (CA_4);
\draw[-{Stealth[scale=1.4,angle'=45]}] (CC_3) to (CC_4);
\draw[-{Stealth[scale=1.4,angle'=45]}] (CC_3) to (CG_4);
\draw[-{Stealth[scale=1.4,angle'=45]}] (CC_3) to (CT_4);
\draw[-{Stealth[scale=1.4,angle'=45]}] (CG_3) to (GA_4);
\draw[-{Stealth[scale=1.4,angle'=45]}] (CG_3) to (GC_4);
\draw[-{Stealth[scale=1.4,angle'=45]}] (CG_3) to (GG_4);
\draw[-{Stealth[scale=1.4,angle'=45]}] (CG_3) to (GT_4);
\draw[-{Stealth[scale=1.4,angle'=45]}] (CT_3) to (TA_4);
\draw[-{Stealth[scale=1.4,angle'=45]}] (CT_3) to (TC_4);
\draw[-{Stealth[scale=1.4,angle'=45]}] (CT_3) to (TG_4);
\draw[-{Stealth[scale=1.4,angle'=45]}] (CT_3) to (TT_4);
\draw[-{Stealth[scale=1.4,angle'=45]}] (GA_3) to (AA_4);
\draw[-{Stealth[scale=1.4,angle'=45]}] (GA_3) to (AC_4);
\draw[-{Stealth[scale=1.4,angle'=45]}] (GA_3) to (AG_4);
\draw[-{Stealth[scale=1.4,angle'=45]}] (GA_3) to (AT_4);
\draw[-{Stealth[scale=1.4,angle'=45]}] (GC_3) to (CA_4);
\draw[-{Stealth[scale=1.4,angle'=45]}] (GC_3) to (CC_4);
\draw[-{Stealth[scale=1.4,angle'=45]}] (GC_3) to (CG_4);
\draw[-{Stealth[scale=1.4,angle'=45]}] (GC_3) to (CT_4);
\draw[-{Stealth[scale=1.4,angle'=45]}] (GG_3) to (GA_4);
\draw[-{Stealth[scale=1.4,angle'=45]}] (GG_3) to (GC_4);
\draw[-{Stealth[scale=1.4,angle'=45]}] (GG_3) to (GG_4);
\draw[-{Stealth[scale=1.4,angle'=45]}] (GG_3) to (GT_4);
\draw[-{Stealth[scale=1.4,angle'=45]}] (GT_3) to (TA_4);
\draw[-{Stealth[scale=1.4,angle'=45]}] (GT_3) to (TC_4);
\draw[-{Stealth[scale=1.4,angle'=45]}] (GT_3) to (TG_4);
\draw[-{Stealth[scale=1.4,angle'=45]}] (GT_3) to (TT_4);
\draw[-{Stealth[scale=1.4,angle'=45]}] (TA_3) to (AA_4);
\draw[-{Stealth[scale=1.4,angle'=45]}] (TA_3) to (AC_4);
\draw[-{Stealth[scale=1.4,angle'=45]}, ultra thick] (TA_3) to (AG_4);
\draw[-{Stealth[scale=1.4,angle'=45]}] (TA_3) to (AT_4);
\draw[-{Stealth[scale=1.4,angle'=45]}] (TC_3) to (CA_4);
\draw[-{Stealth[scale=1.4,angle'=45]}] (TC_3) to (CC_4);
\draw[-{Stealth[scale=1.4,angle'=45]}] (TC_3) to (CG_4);
\draw[-{Stealth[scale=1.4,angle'=45]}] (TC_3) to (CT_4);
\draw[-{Stealth[scale=1.4,angle'=45]}] (TG_3) to (GA_4);
\draw[-{Stealth[scale=1.4,angle'=45]}] (TG_3) to (GC_4);
\draw[-{Stealth[scale=1.4,angle'=45]}] (TG_3) to (GG_4);
\draw[-{Stealth[scale=1.4,angle'=45]}] (TG_3) to (GT_4);
\draw[-{Stealth[scale=1.4,angle'=45]}] (TT_3) to (TA_4);
\draw[-{Stealth[scale=1.4,angle'=45]}] (TT_3) to (TC_4);
\draw[-{Stealth[scale=1.4,angle'=45]}] (TT_3) to (TG_4);
\draw[-{Stealth[scale=1.4,angle'=45]}] (TT_3) to (TT_4);
\draw[-{Stealth[scale=1.4,angle'=45]}] (AA_4) to (AA_5);
\draw[-{Stealth[scale=1.4,angle'=45]}] (AA_4) to (AC_5);
\draw[-{Stealth[scale=1.4,angle'=45]}] (AA_4) to (AG_5);
\draw[-{Stealth[scale=1.4,angle'=45]}] (AA_4) to (AT_5);
\draw[-{Stealth[scale=1.4,angle'=45]}] (AC_4) to (CA_5);
\draw[-{Stealth[scale=1.4,angle'=45]}] (AC_4) to (CC_5);
\draw[-{Stealth[scale=1.4,angle'=45]}] (AC_4) to (CG_5);
\draw[-{Stealth[scale=1.4,angle'=45]}] (AC_4) to (CT_5);
\draw[-{Stealth[scale=1.4,angle'=45]}, ultra thick] (AG_4) to (GA_5);
\draw[-{Stealth[scale=1.4,angle'=45]}] (AG_4) to (GC_5);
\draw[-{Stealth[scale=1.4,angle'=45]}] (AG_4) to (GG_5);
\draw[-{Stealth[scale=1.4,angle'=45]}] (AG_4) to (GT_5);
\draw[-{Stealth[scale=1.4,angle'=45]}] (AT_4) to (TA_5);
\draw[-{Stealth[scale=1.4,angle'=45]}] (AT_4) to (TC_5);
\draw[-{Stealth[scale=1.4,angle'=45]}] (AT_4) to (TG_5);
\draw[-{Stealth[scale=1.4,angle'=45]}] (AT_4) to (TT_5);
\draw[-{Stealth[scale=1.4,angle'=45]}] (CA_4) to (AA_5);
\draw[-{Stealth[scale=1.4,angle'=45]}] (CA_4) to (AC_5);
\draw[-{Stealth[scale=1.4,angle'=45]}] (CA_4) to (AG_5);
\draw[-{Stealth[scale=1.4,angle'=45]}] (CA_4) to (AT_5);
\draw[-{Stealth[scale=1.4,angle'=45]}] (CC_4) to (CA_5);
\draw[-{Stealth[scale=1.4,angle'=45]}] (CC_4) to (CC_5);
\draw[-{Stealth[scale=1.4,angle'=45]}] (CC_4) to (CG_5);
\draw[-{Stealth[scale=1.4,angle'=45]}] (CC_4) to (CT_5);
\draw[-{Stealth[scale=1.4,angle'=45]}] (CG_4) to (GA_5);
\draw[-{Stealth[scale=1.4,angle'=45]}] (CG_4) to (GC_5);
\draw[-{Stealth[scale=1.4,angle'=45]}] (CG_4) to (GG_5);
\draw[-{Stealth[scale=1.4,angle'=45]}] (CG_4) to (GT_5);
\draw[-{Stealth[scale=1.4,angle'=45]}] (CT_4) to (TA_5);
\draw[-{Stealth[scale=1.4,angle'=45]}] (CT_4) to (TC_5);
\draw[-{Stealth[scale=1.4,angle'=45]}] (CT_4) to (TG_5);
\draw[-{Stealth[scale=1.4,angle'=45]}] (CT_4) to (TT_5);
\draw[-{Stealth[scale=1.4,angle'=45]}] (GA_4) to (AA_5);
\draw[-{Stealth[scale=1.4,angle'=45]}] (GA_4) to (AC_5);
\draw[-{Stealth[scale=1.4,angle'=45]}] (GA_4) to (AG_5);
\draw[-{Stealth[scale=1.4,angle'=45]}] (GA_4) to (AT_5);
\draw[-{Stealth[scale=1.4,angle'=45]}] (GC_4) to (CA_5);
\draw[-{Stealth[scale=1.4,angle'=45]}] (GC_4) to (CC_5);
\draw[-{Stealth[scale=1.4,angle'=45]}] (GC_4) to (CG_5);
\draw[-{Stealth[scale=1.4,angle'=45]}] (GC_4) to (CT_5);
\draw[-{Stealth[scale=1.4,angle'=45]}] (GG_4) to (GA_5);
\draw[-{Stealth[scale=1.4,angle'=45]}] (GG_4) to (GC_5);
\draw[-{Stealth[scale=1.4,angle'=45]}] (GG_4) to (GG_5);
\draw[-{Stealth[scale=1.4,angle'=45]}] (GG_4) to (GT_5);
\draw[-{Stealth[scale=1.4,angle'=45]}] (GT_4) to (TA_5);
\draw[-{Stealth[scale=1.4,angle'=45]}] (GT_4) to (TC_5);
\draw[-{Stealth[scale=1.4,angle'=45]}] (GT_4) to (TG_5);
\draw[-{Stealth[scale=1.4,angle'=45]}] (GT_4) to (TT_5);
\draw[-{Stealth[scale=1.4,angle'=45]}] (TA_4) to (AA_5);
\draw[-{Stealth[scale=1.4,angle'=45]}] (TA_4) to (AC_5);
\draw[-{Stealth[scale=1.4,angle'=45]}] (TA_4) to (AG_5);
\draw[-{Stealth[scale=1.4,angle'=45]}] (TA_4) to (AT_5);
\draw[-{Stealth[scale=1.4,angle'=45]}] (TC_4) to (CA_5);
\draw[-{Stealth[scale=1.4,angle'=45]}] (TC_4) to (CC_5);
\draw[-{Stealth[scale=1.4,angle'=45]}] (TC_4) to (CG_5);
\draw[-{Stealth[scale=1.4,angle'=45]}] (TC_4) to (CT_5);
\draw[-{Stealth[scale=1.4,angle'=45]}] (TG_4) to (GA_5);
\draw[-{Stealth[scale=1.4,angle'=45]}] (TG_4) to (GC_5);
\draw[-{Stealth[scale=1.4,angle'=45]}] (TG_4) to (GG_5);
\draw[-{Stealth[scale=1.4,angle'=45]}] (TG_4) to (GT_5);
\draw[-{Stealth[scale=1.4,angle'=45]}] (TT_4) to (TA_5);
\draw[-{Stealth[scale=1.4,angle'=45]}] (TT_4) to (TC_5);
\draw[-{Stealth[scale=1.4,angle'=45]}] (TT_4) to (TG_5);
\draw[-{Stealth[scale=1.4,angle'=45]}] (TT_4) to (TT_5);
\end{tikzpicture}}
\caption{The graph $B_{6,3}$. The path defined by the sequence \texttt{AATAGA} is highlighted.\label{fig:debruijn}}
\end{center}
\end{figure}

%% file: fig/fig_greedy.tex
\begin{tikzpicture}[scale=0.9, /tikz/background rectangle/.style={fill={rgb,1:red,1.0;green,1.0;blue,1.0}, draw opacity={1.0}}, show background rectangle]
    \begin{axis}[point meta max={nan}, point meta min={nan}, legend cell align={left}, title={}, title style={at={{(0.5,1)}}, anchor={south}, font={{\fontsize{14 pt}{18.2 pt}\selectfont}}, color={rgb,1:red,0.0;green,0.0;blue,0.0}, draw opacity={1.0}, rotate={0.0}}, legend style={color={rgb,1:red,0.0;green,0.0;blue,0.0}, draw opacity={1.0}, line width={1}, solid, fill={rgb,1:red,1.0;green,1.0;blue,1.0}, fill opacity={1.0}, text opacity={1.0}, font={{\fontsize{8 pt}{10.4 pt}\selectfont}}, text={rgb,1:red,0.0;green,0.0;blue,0.0}, at={(0.98, 0.02)}, anchor={south east}}, axis background/.style={fill={rgb,1:red,1.0;green,1.0;blue,1.0}, opacity={1.0}}, anchor={north west}, xshift={1.0mm}, yshift={-1.0mm}, width={99.6mm}, height={74.2mm}, scaled x ticks={false}, xlabel={$\textrm{Time (seconds)}$}, x tick style={color={rgb,1:red,0.0;green,0.0;blue,0.0}, opacity={1.0}}, x tick label style={color={rgb,1:red,0.0;green,0.0;blue,0.0}, opacity={1.0}, rotate={0}}, xlabel style={at={(ticklabel cs:0.5)}, anchor=near ticklabel, font={{\fontsize{11 pt}{14.3 pt}\selectfont}}, color={rgb,1:red,0.0;green,0.0;blue,0.0}, draw opacity={1.0}, rotate={0.0}}, xmajorgrids={false}, xmin={-3.4703790822300005}, xmax={119.14968182323001}, xtick={{0.0,25.0,50.0,75.0,100.0}}, xticklabels={{$0$,$25$,$50$,$75$,$100$}}, xtick align={inside}, xticklabel style={font={{\fontsize{8 pt}{10.4 pt}\selectfont}}, color={rgb,1:red,0.0;green,0.0;blue,0.0}, draw opacity={1.0}, rotate={0.0}}, x grid style={color={rgb,1:red,0.0;green,0.0;blue,0.0}, draw opacity={0.1}, line width={0.5}, solid}, axis x line*={left}, x axis line style={color={rgb,1:red,0.0;green,0.0;blue,0.0}, draw opacity={1.0}, line width={1}, solid}, scaled y ticks={false}, ylabel={$\ell$}, y tick style={color={rgb,1:red,0.0;green,0.0;blue,0.0}, opacity={1.0}}, y tick label style={color={rgb,1:red,0.0;green,0.0;blue,0.0}, opacity={1.0}, rotate={0}}, ylabel style={at={(ticklabel cs:0.5)}, anchor=near ticklabel, font={{\fontsize{11 pt}{14.3 pt}\selectfont}}, color={rgb,1:red,0.0;green,0.0;blue,0.0}, draw opacity={1.0}, rotate={0.0}}, ymajorgrids={false}, ymin={-7425.540469622827}, ymax={-3224.5444742572936}, ytick={{-7000.0,-6000.0,-5000.0,-4000.0}}, yticklabels={{$-7000$,$-6000$,$-5000$,$-4000$}}, ytick align={inside}, yticklabel style={font={{\fontsize{8 pt}{10.4 pt}\selectfont}}, color={rgb,1:red,0.0;green,0.0;blue,0.0}, draw opacity={1.0}, rotate={0.0}}, y grid style={color={rgb,1:red,0.0;green,0.0;blue,0.0}, draw opacity={0.1}, line width={0.5}, solid}, axis y line*={left}, y axis line style={color={rgb,1:red,0.0;green,0.0;blue,0.0}, draw opacity={1.0}, line width={1}, solid}]
        \addplot[color={rgb,1:red,0.0;green,0.6056;blue,0.9787}, name path={ff064312-f558-4861-825f-33e834b2e43b}, draw opacity={1.0}, line width={1}, solid]
            table[row sep={\\}]
            {
                \\
                0.0  -6486.945794802212  \\
                0.006643324000000001  -4208.831387050738  \\
                0.01810788  -4208.831387050738  \\
                0.029130524  -4208.831387050738  \\
                0.041134216999999994  -4208.831387050738  \\
                0.049260668  -4208.831387050738  \\
                0.060608327999999996  -4208.831387050738  \\
                0.072904404  -4208.831387050738  \\
                0.08425806999999999  -4208.831387050738  \\
                0.09359821  -4208.831387050738  \\
                0.10504659699999999  -4208.831387050738  \\
                0.11674736699999999  -4068.1355468696033  \\
                0.12646212799999998  -4068.1355468696033  \\
                0.788550721  -4068.1355468696033  \\
                0.8004213489999998  -4068.1355468696033  \\
                0.8114524409999999  -3978.1788666252523  \\
                0.8253780409999999  -3978.1788666252523  \\
                0.8376059619999998  -3978.1788666252523  \\
                0.8509449179999997  -3978.1788666252523  \\
                0.8616792369999997  -3978.1788666252523  \\
                1.5273188679999998  -3978.1788666252523  \\
                1.538754098  -3978.1788666252523  \\
                1.552668004  -3796.0557738529487  \\
                1.563144688  -3796.0557738529487  \\
                2.2402730459999995  -3796.0557738529487  \\
                2.2529701179999995  -3796.0557738529487  \\
                2.267349678  -3796.0557738529487  \\
                2.277255682  -3796.0557738529487  \\
                2.936687985  -3796.0557738529487  \\
                2.9499360909999996  -3343.440587333677  \\
                2.960421419  -3343.440587333677  \\
                2.9739746039999995  -3343.440587333677  \\
                3.654794576999999  -3343.440587333677  \\
                3.666218309999999  -3343.440587333677  \\
                3.679327226999999  -3343.440587333677  \\
                3.691796263999999  -3343.440587333677  \\
                4.351694219999999  -3343.440587333677  \\
                4.363335944999999  -3343.440587333677  \\
                4.374493568999999  -3343.440587333677  \\
                4.387361489999999  -3343.440587333677  \\
                5.073157433999999  -3343.440587333677  \\
                5.084893916999999  -3343.440587333677  \\
                5.096708662999999  -3343.440587333677  \\
                5.109320024999999  -3343.440587333677  \\
                5.767611883  -3343.440587333677  \\
                5.780384908  -3343.440587333677  \\
                5.792782579  -3343.440587333677  \\
                5.803146362  -3343.440587333677  \\
                6.486910736  -3343.440587333677  \\
                6.498422133  -3343.440587333677  \\
            }
            ;
        \addlegendentry {$\textrm{CGM}$}
        \addplot[color={rgb,1:red,0.8889;green,0.4356;blue,0.2781}, name path={0666711b-da6d-4edc-a7a8-fc2c8ecd0fe8}, draw opacity={1.0}, line width={1}, dashed]
            table[row sep={\\}]
            {
                \\
                1.487363213  -7306.644356546444  \\
                2.9712506640000003  -7152.121798276137  \\
                4.35159597  -7017.288030258852  \\
                5.724220787  -6895.212066596873  \\
                7.0990558870000005  -6793.73757072847  \\
                9.012339974  -6701.326487511501  \\
                10.497382269  -6620.525367018505  \\
                11.959759911999999  -6544.200166760775  \\
                13.430423830999999  -6469.548151148323  \\
                15.033148672  -6397.506818218377  \\
                16.816102611999998  -6328.5403392925  \\
                18.261288489  -6256.383477947234  \\
                19.669189063999998  -6192.022581363311  \\
                20.971254714999997  -6136.250560840831  \\
                22.268690889  -6085.015471870572  \\
                23.560049191999997  -6034.733734533633  \\
                24.998589182999996  -5954.826504678753  \\
                26.425657428999997  -5898.316514059192  \\
                27.857041393999996  -5848.0959585632845  \\
                29.288916990999997  -5809.156852706779  \\
                30.790155675999998  -5771.197594708847  \\
                32.286965503  -5733.517464176936  \\
                33.68075843  -5695.718982620749  \\
                35.099273466999996  -5660.748143008373  \\
                36.429915634  -5602.346416802632  \\
                37.685208163  -5570.217205226138  \\
                38.909931799  -5544.601460416812  \\
                40.127438919  -5490.272602997069  \\
                41.613282343  -5461.637192744858  \\
                43.106603167  -5435.903227203804  \\
                44.608174239  -5410.619781889639  \\
                46.097013017  -5388.5849517120605  \\
                47.578101148  -5361.966377336277  \\
                49.058822755  -5334.892845615111  \\
                50.541327348  -5306.676343703918  \\
                51.785481789  -5226.057756750564  \\
                53.146236309  -5164.96947338318  \\
                54.595981875  -5131.857998847217  \\
                55.73326133  -5097.869512757107  \\
                57.058346129  -5070.8193666493025  \\
                58.51705817  -5047.482944167875  \\
                59.970771487  -5025.631651150644  \\
                61.436066508  -5003.768779709607  \\
                62.881951897  -4984.940165385138  \\
                64.307306331  -4968.603307865472  \\
                65.71897618300001  -4947.162348453576  \\
                67.068895006  -4897.499427580347  \\
                68.192215137  -4852.258473352435  \\
                69.528634376  -4843.4463900641595  \\
                71.04576659  -4835.158571128688  \\
                72.1739828  -4767.231328797829  \\
                73.528783363  -4739.67199956325  \\
                74.999643567  -4734.375853894371  \\
                76.463593599  -4729.925857257954  \\
                77.835967537  -4717.190085540575  \\
                79.210005571  -4714.150600222772  \\
                80.584139972  -4702.580234779737  \\
                82.030706318  -4642.557687842265  \\
                83.305332717  -4621.496926258609  \\
                84.488494477  -4601.214413463648  \\
                85.672688491  -4598.834996169612  \\
                86.855288795  -4596.877116096974  \\
                88.037547054  -4576.246668236168  \\
                89.312861254  -4574.420899914607  \\
                90.682683469  -4566.079868350782  \\
                92.053063241  -4555.504621744584  \\
                93.478071766  -4554.540229441551  \\
                95.281986281  -4550.979152117069  \\
                96.617247811  -4549.4693013895985  \\
                97.940677063  -4549.197531136577  \\
                99.15771556  -4548.782573254312  \\
                100.307793777  -4542.188339886467  \\
                101.43359087900001  -4535.28301215555  \\
                102.55559336600001  -4526.305996309688  \\
                103.73904502200001  -4525.505755175734  \\
                104.90356714600001  -4523.836091233185  \\
                106.27562360600001  -4523.5046320526135  \\
                107.65288071900001  -4523.114444716245  \\
                109.095451612  -4522.912380162329  \\
                110.470960399  -4522.844620225222  \\
                111.847590623  -4522.79035669855  \\
                113.21566628000001  -4522.782706575756  \\
                114.55390727900001  -4522.7821726089605  \\
                115.67930274100001  -4522.782106079715  \\
            }
            ;
        \addlegendentry {$\textrm{Greedy}$}
        \addplot[color={rgb,1:red,0.2422;green,0.6433;blue,0.3044}, name path={6623d70f-6e7b-4ec8-9ff5-cee6f032bd2f}, draw opacity={1.0}, line width={1}, dotted]
            table[row sep={\\}]
            {
                \\
                -126.09043998769  -3379.9955780698424  \\
                241.76974272869  -3379.9955780698424  \\
            }
            ;
        \addlegendentry {$\ell(s_*)$}
    \end{axis}
    \end{tikzpicture}
    

%% file: fig/fig_greedy_errors.tex
\begin{tikzpicture}[scale=0.9, /tikz/background rectangle/.style={fill={rgb,1:red,1.0;green,1.0;blue,1.0}, draw opacity={1.0}}, show background rectangle]
\begin{axis}[point meta max={nan}, point meta min={nan}, legend cell align={left}, title={}, title style={at={{(0.5,1)}}, anchor={south}, font={{\fontsize{14 pt}{18.2 pt}\selectfont}}, color={rgb,1:red,0.0;green,0.0;blue,0.0}, draw opacity={1.0}, rotate={0.0}}, legend style={color={rgb,1:red,0.0;green,0.0;blue,0.0}, draw opacity={1.0}, line width={1}, solid, fill={rgb,1:red,1.0;green,1.0;blue,1.0}, fill opacity={1.0}, text opacity={1.0}, font={{\fontsize{8 pt}{10.4 pt}\selectfont}}, text={rgb,1:red,0.0;green,0.0;blue,0.0}, at={(1.02, 1)}, anchor={north west}}, axis background/.style={fill={rgb,1:red,1.0;green,1.0;blue,1.0}, opacity={1.0}}, anchor={north west}, xshift={1.0mm}, yshift={-1.0mm}, width={99.6mm}, height={74.2mm}, scaled x ticks={false}, xlabel={}, x tick style={color={rgb,1:red,0.0;green,0.0;blue,0.0}, opacity={1.0}}, x tick label style={color={rgb,1:red,0.0;green,0.0;blue,0.0}, opacity={1.0}, rotate={0}}, xlabel style={at={(ticklabel cs:0.5)}, anchor=near ticklabel, font={{\fontsize{11 pt}{14.3 pt}\selectfont}}, color={rgb,1:red,0.0;green,0.0;blue,0.0}, draw opacity={1.0}, rotate={0.0}}, xmajorgrids={false}, xmin={-0.011240000000000007}, xmax={2.01124}, xtick={{0.5,1.5}}, xticklabels={{$$\textrm{CGM}$$,$$\textrm{Greedy}$$}}, xtick align={inside}, xticklabel style={font={{\fontsize{8 pt}{10.4 pt}\selectfont}}, color={rgb,1:red,0.0;green,0.0;blue,0.0}, draw opacity={1.0}, rotate={0.0}}, x grid style={color={rgb,1:red,0.0;green,0.0;blue,0.0}, draw opacity={0.1}, line width={0.5}, solid}, axis x line*={left}, x axis line style={color={rgb,1:red,0.0;green,0.0;blue,0.0}, draw opacity={1.0}, line width={1}, solid}, scaled y ticks={false}, ylabel={$\textrm{Error Rate}$}, y tick style={color={rgb,1:red,0.0;green,0.0;blue,0.0}, opacity={1.0}}, y tick label style={color={rgb,1:red,0.0;green,0.0;blue,0.0}, opacity={1.0}, rotate={0}}, ylabel style={at={(ticklabel cs:0.5)}, anchor=near ticklabel, font={{\fontsize{11 pt}{14.3 pt}\selectfont}}, color={rgb,1:red,0.0;green,0.0;blue,0.0}, draw opacity={1.0}, rotate={0.0}}, ymajorgrids={false}, ymin={-0.011286}, ymax={0.387486}, ytick={{0.0,0.1,0.2,0.30000000000000004}}, yticklabels={{$0.0$,$0.1$,$0.2$,$0.3$}}, ytick align={inside}, yticklabel style={font={{\fontsize{8 pt}{10.4 pt}\selectfont}}, color={rgb,1:red,0.0;green,0.0;blue,0.0}, draw opacity={1.0}, rotate={0.0}}, y grid style={color={rgb,1:red,0.0;green,0.0;blue,0.0}, draw opacity={0.1}, line width={0.5}, solid}, axis y line*={left}, y axis line style={color={rgb,1:red,0.0;green,0.0;blue,0.0}, draw opacity={1.0}, line width={1}, solid}]
    \addplot[color={rgb,1:red,0.0;green,0.0;blue,0.0}, name path={8b93ca71-e17a-4cad-8289-f702bef2dcc9}, area legend, fill={rgb,1:red,0.502;green,0.502;blue,0.502}, fill opacity={1.0}, draw opacity={1.0}, line width={1}, solid]
        table[row sep={\\}]
        {
            \\
            0.09999999999999998  0.0004  \\
            0.09999999999999998  0.0  \\
            0.9  0.0  \\
            0.9  0.0004  \\
            0.09999999999999998  0.0004  \\
        }
        ;
    \addplot[color={rgb,1:red,0.0;green,0.0;blue,0.0}, name path={8b93ca71-e17a-4cad-8289-f702bef2dcc9}, area legend, fill={rgb,1:red,0.502;green,0.502;blue,0.502}, fill opacity={1.0}, draw opacity={1.0}, line width={1}, solid]
        table[row sep={\\}]
        {
            \\
            1.1  0.36119999999999997  \\
            1.1  0.0  \\
            1.9  0.0  \\
            1.9  0.36119999999999997  \\
            1.1  0.36119999999999997  \\
        }
        ;
    \addplot[color={rgb,1:red,0.0;green,0.0;blue,0.0}, name path={109e3d79-7ed4-4249-a14c-baffb66cfa91}, draw opacity={1.0}, line width={1}, solid, mark={-}, mark size={3.0 pt}, mark repeat={1}, mark options={color={rgb,1:red,0.0;green,0.0;blue,0.0}, draw opacity={1.0}, fill={rgb,1:red,0.0;green,0.0;blue,0.0}, fill opacity={1.0}, line width={0.75}, rotate={0}, solid}]
        table[row sep={\\}]
        {
            \\
            0.5  0.0  \\
            0.5  0.0012  \\
        }
        ;
    \addplot[color={rgb,1:red,0.0;green,0.0;blue,0.0}, name path={109e3d79-7ed4-4249-a14c-baffb66cfa91}, draw opacity={1.0}, line width={1}, solid, mark={-}, mark size={3.0 pt}, mark repeat={1}, mark options={color={rgb,1:red,0.0;green,0.0;blue,0.0}, draw opacity={1.0}, fill={rgb,1:red,0.0;green,0.0;blue,0.0}, fill opacity={1.0}, line width={0.75}, rotate={0}, solid}]
        table[row sep={\\}]
        {
            \\
            1.5  0.34480000000000005  \\
            1.5  0.3762  \\
        }
        ;
\end{axis}
\end{tikzpicture}

%% file: fig/fig_probes.tex
\begin{tikzpicture}[/tikz/background rectangle/.style={fill={rgb,1:red,1.0;green,1.0;blue,1.0}, draw opacity={1.0}}, show background rectangle]
\begin{axis}[point meta max={nan}, point meta min={nan}, legend cell align={left}, title={}, title style={at={{(0.5,1)}}, anchor={south}, font={{\fontsize{14 pt}{18.2 pt}\selectfont}}, color={rgb,1:red,0.0;green,0.0;blue,0.0}, draw opacity={1.0}, rotate={0.0}}, legend style={color={rgb,1:red,0.0;green,0.0;blue,0.0}, draw opacity={1.0}, line width={1}, solid, fill={rgb,1:red,1.0;green,1.0;blue,1.0}, fill opacity={1.0}, text opacity={1.0}, font={{\fontsize{8 pt}{10.4 pt}\selectfont}}, text={rgb,1:red,0.0;green,0.0;blue,0.0}, at={(1.02, 1)}, anchor={north west}}, axis background/.style={fill={rgb,1:red,1.0;green,1.0;blue,1.0}, opacity={1.0}}, anchor={north west}, xshift={1.0mm}, yshift={-1.0mm}, width={99.6mm}, height={74.2mm}, scaled x ticks={false}, xlabel={$\textrm{Number of Probes}$}, x tick style={color={rgb,1:red,0.0;green,0.0;blue,0.0}, opacity={1.0}}, x tick label style={color={rgb,1:red,0.0;green,0.0;blue,0.0}, opacity={1.0}, rotate={0}}, xlabel style={at={(ticklabel cs:0.5)}, anchor=near ticklabel, font={{\fontsize{11 pt}{14.3 pt}\selectfont}}, color={rgb,1:red,0.0;green,0.0;blue,0.0}, draw opacity={1.0}, rotate={0.0}}, xmajorgrids={false}, xmin={3.37}, xmax={25.63}, xtick={{5.0,10.0,15.0,20.0,25.0}}, xticklabels={{$5$,$10$,$15$,$20$,$25$}}, xtick align={inside}, xticklabel style={font={{\fontsize{8 pt}{10.4 pt}\selectfont}}, color={rgb,1:red,0.0;green,0.0;blue,0.0}, draw opacity={1.0}, rotate={0.0}}, x grid style={color={rgb,1:red,0.0;green,0.0;blue,0.0}, draw opacity={0.1}, line width={0.5}, solid}, axis x line*={left}, x axis line style={color={rgb,1:red,0.0;green,0.0;blue,0.0}, draw opacity={1.0}, line width={1}, solid}, scaled y ticks={false}, ylabel={$\textrm{Error Rate}$}, y tick style={color={rgb,1:red,0.0;green,0.0;blue,0.0}, opacity={1.0}}, y tick label style={color={rgb,1:red,0.0;green,0.0;blue,0.0}, opacity={1.0}, rotate={0}}, ylabel style={at={(ticklabel cs:0.5)}, anchor=near ticklabel, font={{\fontsize{11 pt}{14.3 pt}\selectfont}}, color={rgb,1:red,0.0;green,0.0;blue,0.0}, draw opacity={1.0}, rotate={0.0}}, ymajorgrids={false}, ymin={-0.0007739999999999986}, ymax={0.16437400000000002}, ytick={{0.0,0.05,0.1,0.15000000000000002}}, yticklabels={{$0.00$,$0.05$,$0.10$,$0.15$}}, ytick align={inside}, yticklabel style={font={{\fontsize{8 pt}{10.4 pt}\selectfont}}, color={rgb,1:red,0.0;green,0.0;blue,0.0}, draw opacity={1.0}, rotate={0.0}}, y grid style={color={rgb,1:red,0.0;green,0.0;blue,0.0}, draw opacity={0.1}, line width={0.5}, solid}, axis y line*={left}, y axis line style={color={rgb,1:red,0.0;green,0.0;blue,0.0}, draw opacity={1.0}, line width={1}, solid}]
    \addplot[color={rgb,1:red,0.0;green,0.0;blue,0.0}, name path={32c05278-f4f6-47e7-9dad-72ceeeb798e9}, draw opacity={1.0}, line width={1}, solid]
        table[row sep={\\}]
        {
            \\
            4.0  0.1424  \\
            5.0  0.0757  \\
            6.0  0.05140000000000001  \\
            7.0  0.0219  \\
            8.0  0.0085  \\
            9.0  0.013900000000000001  \\
            10.0  0.0094  \\
            11.0  0.013500000000000002  \\
            12.0  0.0093  \\
            13.0  0.014500000000000002  \\
            14.0  0.0127  \\
            15.0  0.0202  \\
            16.0  0.0187  \\
            17.0  0.017400000000000002  \\
            18.0  0.023000000000000003  \\
            19.0  0.0395  \\
            20.0  0.049699999999999994  \\
            21.0  0.063  \\
            22.0  0.06169999999999999  \\
            23.0  0.0812  \\
            24.0  0.09040000000000001  \\
            25.0  0.13060000000000002  \\
        }
        ;
    \addplot[color={rgb,1:red,0.0;green,0.0;blue,0.0}, name path={ea8cb4f4-5805-4109-9862-120d61a8b5da}, draw opacity={1.0}, line width={1}, solid, mark={-}, mark size={3.0 pt}, mark repeat={1}, mark options={color={rgb,1:red,0.0;green,0.0;blue,0.0}, draw opacity={1.0}, fill={rgb,1:red,0.0;green,0.0;blue,0.0}, fill opacity={1.0}, line width={0.75}, rotate={0}, solid}]
        table[row sep={\\}]
        {
            \\
            4.0  0.12519750000000002  \\
            4.0  0.1597  \\
        }
        ;
    \addplot[color={rgb,1:red,0.0;green,0.0;blue,0.0}, name path={ea8cb4f4-5805-4109-9862-120d61a8b5da}, draw opacity={1.0}, line width={1}, solid, mark={-}, mark size={3.0 pt}, mark repeat={1}, mark options={color={rgb,1:red,0.0;green,0.0;blue,0.0}, draw opacity={1.0}, fill={rgb,1:red,0.0;green,0.0;blue,0.0}, fill opacity={1.0}, line width={0.75}, rotate={0}, solid}]
        table[row sep={\\}]
        {
            \\
            5.0  0.06349500000000001  \\
            5.0  0.08870249999999999  \\
        }
        ;
    \addplot[color={rgb,1:red,0.0;green,0.0;blue,0.0}, name path={ea8cb4f4-5805-4109-9862-120d61a8b5da}, draw opacity={1.0}, line width={1}, solid, mark={-}, mark size={3.0 pt}, mark repeat={1}, mark options={color={rgb,1:red,0.0;green,0.0;blue,0.0}, draw opacity={1.0}, fill={rgb,1:red,0.0;green,0.0;blue,0.0}, fill opacity={1.0}, line width={0.75}, rotate={0}, solid}]
        table[row sep={\\}]
        {
            \\
            6.0  0.0405975  \\
            6.0  0.0635025  \\
        }
        ;
    \addplot[color={rgb,1:red,0.0;green,0.0;blue,0.0}, name path={ea8cb4f4-5805-4109-9862-120d61a8b5da}, draw opacity={1.0}, line width={1}, solid, mark={-}, mark size={3.0 pt}, mark repeat={1}, mark options={color={rgb,1:red,0.0;green,0.0;blue,0.0}, draw opacity={1.0}, fill={rgb,1:red,0.0;green,0.0;blue,0.0}, fill opacity={1.0}, line width={0.75}, rotate={0}, solid}]
        table[row sep={\\}]
        {
            \\
            7.0  0.014400000000000001  \\
            7.0  0.029500000000000002  \\
        }
        ;
    \addplot[color={rgb,1:red,0.0;green,0.0;blue,0.0}, name path={ea8cb4f4-5805-4109-9862-120d61a8b5da}, draw opacity={1.0}, line width={1}, solid, mark={-}, mark size={3.0 pt}, mark repeat={1}, mark options={color={rgb,1:red,0.0;green,0.0;blue,0.0}, draw opacity={1.0}, fill={rgb,1:red,0.0;green,0.0;blue,0.0}, fill opacity={1.0}, line width={0.75}, rotate={0}, solid}]
        table[row sep={\\}]
        {
            \\
            8.0  0.0039000000000000007  \\
            8.0  0.014499999999999999  \\
        }
        ;
    \addplot[color={rgb,1:red,0.0;green,0.0;blue,0.0}, name path={ea8cb4f4-5805-4109-9862-120d61a8b5da}, draw opacity={1.0}, line width={1}, solid, mark={-}, mark size={3.0 pt}, mark repeat={1}, mark options={color={rgb,1:red,0.0;green,0.0;blue,0.0}, draw opacity={1.0}, fill={rgb,1:red,0.0;green,0.0;blue,0.0}, fill opacity={1.0}, line width={0.75}, rotate={0}, solid}]
        table[row sep={\\}]
        {
            \\
            9.0  0.008397500000000002  \\
            9.0  0.0199025  \\
        }
        ;
    \addplot[color={rgb,1:red,0.0;green,0.0;blue,0.0}, name path={ea8cb4f4-5805-4109-9862-120d61a8b5da}, draw opacity={1.0}, line width={1}, solid, mark={-}, mark size={3.0 pt}, mark repeat={1}, mark options={color={rgb,1:red,0.0;green,0.0;blue,0.0}, draw opacity={1.0}, fill={rgb,1:red,0.0;green,0.0;blue,0.0}, fill opacity={1.0}, line width={0.75}, rotate={0}, solid}]
        table[row sep={\\}]
        {
            \\
            10.0  0.004797500000000002  \\
            10.0  0.015499999999999998  \\
        }
        ;
    \addplot[color={rgb,1:red,0.0;green,0.0;blue,0.0}, name path={ea8cb4f4-5805-4109-9862-120d61a8b5da}, draw opacity={1.0}, line width={1}, solid, mark={-}, mark size={3.0 pt}, mark repeat={1}, mark options={color={rgb,1:red,0.0;green,0.0;blue,0.0}, draw opacity={1.0}, fill={rgb,1:red,0.0;green,0.0;blue,0.0}, fill opacity={1.0}, line width={0.75}, rotate={0}, solid}]
        table[row sep={\\}]
        {
            \\
            11.0  0.0075  \\
            11.0  0.0204  \\
        }
        ;
    \addplot[color={rgb,1:red,0.0;green,0.0;blue,0.0}, name path={ea8cb4f4-5805-4109-9862-120d61a8b5da}, draw opacity={1.0}, line width={1}, solid, mark={-}, mark size={3.0 pt}, mark repeat={1}, mark options={color={rgb,1:red,0.0;green,0.0;blue,0.0}, draw opacity={1.0}, fill={rgb,1:red,0.0;green,0.0;blue,0.0}, fill opacity={1.0}, line width={0.75}, rotate={0}, solid}]
        table[row sep={\\}]
        {
            \\
            12.0  0.0052  \\
            12.0  0.014300000000000002  \\
        }
        ;
    \addplot[color={rgb,1:red,0.0;green,0.0;blue,0.0}, name path={ea8cb4f4-5805-4109-9862-120d61a8b5da}, draw opacity={1.0}, line width={1}, solid, mark={-}, mark size={3.0 pt}, mark repeat={1}, mark options={color={rgb,1:red,0.0;green,0.0;blue,0.0}, draw opacity={1.0}, fill={rgb,1:red,0.0;green,0.0;blue,0.0}, fill opacity={1.0}, line width={0.75}, rotate={0}, solid}]
        table[row sep={\\}]
        {
            \\
            13.0  0.008700000000000001  \\
            13.0  0.0212025  \\
        }
        ;
    \addplot[color={rgb,1:red,0.0;green,0.0;blue,0.0}, name path={ea8cb4f4-5805-4109-9862-120d61a8b5da}, draw opacity={1.0}, line width={1}, solid, mark={-}, mark size={3.0 pt}, mark repeat={1}, mark options={color={rgb,1:red,0.0;green,0.0;blue,0.0}, draw opacity={1.0}, fill={rgb,1:red,0.0;green,0.0;blue,0.0}, fill opacity={1.0}, line width={0.75}, rotate={0}, solid}]
        table[row sep={\\}]
        {
            \\
            14.0  0.008097500000000004  \\
            14.0  0.017902500000000002  \\
        }
        ;
    \addplot[color={rgb,1:red,0.0;green,0.0;blue,0.0}, name path={ea8cb4f4-5805-4109-9862-120d61a8b5da}, draw opacity={1.0}, line width={1}, solid, mark={-}, mark size={3.0 pt}, mark repeat={1}, mark options={color={rgb,1:red,0.0;green,0.0;blue,0.0}, draw opacity={1.0}, fill={rgb,1:red,0.0;green,0.0;blue,0.0}, fill opacity={1.0}, line width={0.75}, rotate={0}, solid}]
        table[row sep={\\}]
        {
            \\
            15.0  0.0127  \\
            15.0  0.0289025  \\
        }
        ;
    \addplot[color={rgb,1:red,0.0;green,0.0;blue,0.0}, name path={ea8cb4f4-5805-4109-9862-120d61a8b5da}, draw opacity={1.0}, line width={1}, solid, mark={-}, mark size={3.0 pt}, mark repeat={1}, mark options={color={rgb,1:red,0.0;green,0.0;blue,0.0}, draw opacity={1.0}, fill={rgb,1:red,0.0;green,0.0;blue,0.0}, fill opacity={1.0}, line width={0.75}, rotate={0}, solid}]
        table[row sep={\\}]
        {
            \\
            16.0  0.0125  \\
            16.0  0.0255025  \\
        }
        ;
    \addplot[color={rgb,1:red,0.0;green,0.0;blue,0.0}, name path={ea8cb4f4-5805-4109-9862-120d61a8b5da}, draw opacity={1.0}, line width={1}, solid, mark={-}, mark size={3.0 pt}, mark repeat={1}, mark options={color={rgb,1:red,0.0;green,0.0;blue,0.0}, draw opacity={1.0}, fill={rgb,1:red,0.0;green,0.0;blue,0.0}, fill opacity={1.0}, line width={0.75}, rotate={0}, solid}]
        table[row sep={\\}]
        {
            \\
            17.0  0.0114  \\
            17.0  0.023700000000000002  \\
        }
        ;
    \addplot[color={rgb,1:red,0.0;green,0.0;blue,0.0}, name path={ea8cb4f4-5805-4109-9862-120d61a8b5da}, draw opacity={1.0}, line width={1}, solid, mark={-}, mark size={3.0 pt}, mark repeat={1}, mark options={color={rgb,1:red,0.0;green,0.0;blue,0.0}, draw opacity={1.0}, fill={rgb,1:red,0.0;green,0.0;blue,0.0}, fill opacity={1.0}, line width={0.75}, rotate={0}, solid}]
        table[row sep={\\}]
        {
            \\
            18.0  0.016800000000000002  \\
            18.0  0.0295025  \\
        }
        ;
    \addplot[color={rgb,1:red,0.0;green,0.0;blue,0.0}, name path={ea8cb4f4-5805-4109-9862-120d61a8b5da}, draw opacity={1.0}, line width={1}, solid, mark={-}, mark size={3.0 pt}, mark repeat={1}, mark options={color={rgb,1:red,0.0;green,0.0;blue,0.0}, draw opacity={1.0}, fill={rgb,1:red,0.0;green,0.0;blue,0.0}, fill opacity={1.0}, line width={0.75}, rotate={0}, solid}]
        table[row sep={\\}]
        {
            \\
            19.0  0.027792500000000005  \\
            19.0  0.05130249999999999  \\
        }
        ;
    \addplot[color={rgb,1:red,0.0;green,0.0;blue,0.0}, name path={ea8cb4f4-5805-4109-9862-120d61a8b5da}, draw opacity={1.0}, line width={1}, solid, mark={-}, mark size={3.0 pt}, mark repeat={1}, mark options={color={rgb,1:red,0.0;green,0.0;blue,0.0}, draw opacity={1.0}, fill={rgb,1:red,0.0;green,0.0;blue,0.0}, fill opacity={1.0}, line width={0.75}, rotate={0}, solid}]
        table[row sep={\\}]
        {
            \\
            20.0  0.038497500000000004  \\
            20.0  0.06169999999999999  \\
        }
        ;
    \addplot[color={rgb,1:red,0.0;green,0.0;blue,0.0}, name path={ea8cb4f4-5805-4109-9862-120d61a8b5da}, draw opacity={1.0}, line width={1}, solid, mark={-}, mark size={3.0 pt}, mark repeat={1}, mark options={color={rgb,1:red,0.0;green,0.0;blue,0.0}, draw opacity={1.0}, fill={rgb,1:red,0.0;green,0.0;blue,0.0}, fill opacity={1.0}, line width={0.75}, rotate={0}, solid}]
        table[row sep={\\}]
        {
            \\
            21.0  0.05119750000000001  \\
            21.0  0.07610250000000002  \\
        }
        ;
    \addplot[color={rgb,1:red,0.0;green,0.0;blue,0.0}, name path={ea8cb4f4-5805-4109-9862-120d61a8b5da}, draw opacity={1.0}, line width={1}, solid, mark={-}, mark size={3.0 pt}, mark repeat={1}, mark options={color={rgb,1:red,0.0;green,0.0;blue,0.0}, draw opacity={1.0}, fill={rgb,1:red,0.0;green,0.0;blue,0.0}, fill opacity={1.0}, line width={0.75}, rotate={0}, solid}]
        table[row sep={\\}]
        {
            \\
            22.0  0.04829749999999999  \\
            22.0  0.07430250000000001  \\
        }
        ;
    \addplot[color={rgb,1:red,0.0;green,0.0;blue,0.0}, name path={ea8cb4f4-5805-4109-9862-120d61a8b5da}, draw opacity={1.0}, line width={1}, solid, mark={-}, mark size={3.0 pt}, mark repeat={1}, mark options={color={rgb,1:red,0.0;green,0.0;blue,0.0}, draw opacity={1.0}, fill={rgb,1:red,0.0;green,0.0;blue,0.0}, fill opacity={1.0}, line width={0.75}, rotate={0}, solid}]
        table[row sep={\\}]
        {
            \\
            23.0  0.0639975  \\
            23.0  0.100505  \\
        }
        ;
    \addplot[color={rgb,1:red,0.0;green,0.0;blue,0.0}, name path={ea8cb4f4-5805-4109-9862-120d61a8b5da}, draw opacity={1.0}, line width={1}, solid, mark={-}, mark size={3.0 pt}, mark repeat={1}, mark options={color={rgb,1:red,0.0;green,0.0;blue,0.0}, draw opacity={1.0}, fill={rgb,1:red,0.0;green,0.0;blue,0.0}, fill opacity={1.0}, line width={0.75}, rotate={0}, solid}]
        table[row sep={\\}]
        {
            \\
            24.0  0.07629749999999999  \\
            24.0  0.1059  \\
        }
        ;
    \addplot[color={rgb,1:red,0.0;green,0.0;blue,0.0}, name path={ea8cb4f4-5805-4109-9862-120d61a8b5da}, draw opacity={1.0}, line width={1}, solid, mark={-}, mark size={3.0 pt}, mark repeat={1}, mark options={color={rgb,1:red,0.0;green,0.0;blue,0.0}, draw opacity={1.0}, fill={rgb,1:red,0.0;green,0.0;blue,0.0}, fill opacity={1.0}, line width={0.75}, rotate={0}, solid}]
        table[row sep={\\}]
        {
            \\
            25.0  0.11279000000000002  \\
            25.0  0.14959999999999998  \\
        }
        ;
\end{axis}
\end{tikzpicture}

%% file: fig/fig_rounds.tex
\begin{tikzpicture}[/tikz/background rectangle/.style={fill={rgb,1:red,1.0;green,1.0;blue,1.0}, draw opacity={1.0}}, show background rectangle]
\begin{axis}[point meta max={nan}, point meta min={nan}, legend cell align={left}, title={}, title style={at={{(0.5,1)}}, anchor={south}, font={{\fontsize{14 pt}{18.2 pt}\selectfont}}, color={rgb,1:red,0.0;green,0.0;blue,0.0}, draw opacity={1.0}, rotate={0.0}}, legend style={color={rgb,1:red,0.0;green,0.0;blue,0.0}, draw opacity={1.0}, line width={1}, solid, fill={rgb,1:red,1.0;green,1.0;blue,1.0}, fill opacity={1.0}, text opacity={1.0}, font={{\fontsize{8 pt}{10.4 pt}\selectfont}}, text={rgb,1:red,0.0;green,0.0;blue,0.0}, at={(1.02, 1)}, anchor={north west}}, axis background/.style={fill={rgb,1:red,1.0;green,1.0;blue,1.0}, opacity={1.0}}, anchor={north west}, xshift={1.0mm}, yshift={-1.0mm}, width={99.6mm}, height={74.2mm}, scaled x ticks={false}, xlabel={$\textrm{Imaging Rounds}$}, x tick style={color={rgb,1:red,0.0;green,0.0;blue,0.0}, opacity={1.0}}, x tick label style={color={rgb,1:red,0.0;green,0.0;blue,0.0}, opacity={1.0}, rotate={0}}, xlabel style={at={(ticklabel cs:0.5)}, anchor=near ticklabel, font={{\fontsize{11 pt}{14.3 pt}\selectfont}}, color={rgb,1:red,0.0;green,0.0;blue,0.0}, draw opacity={1.0}, rotate={0.0}}, xmajorgrids={false}, xmin={96.4}, xmax={223.6}, xtick={{100.0,125.0,150.0,175.0,200.0}}, xticklabels={{$100$,$125$,$150$,$175$,$200$}}, xtick align={inside}, xticklabel style={font={{\fontsize{8 pt}{10.4 pt}\selectfont}}, color={rgb,1:red,0.0;green,0.0;blue,0.0}, draw opacity={1.0}, rotate={0.0}}, x grid style={color={rgb,1:red,0.0;green,0.0;blue,0.0}, draw opacity={0.1}, line width={0.5}, solid}, axis x line*={left}, x axis line style={color={rgb,1:red,0.0;green,0.0;blue,0.0}, draw opacity={1.0}, line width={1}, solid}, scaled y ticks={false}, ylabel={$\textrm{Error Rate}$}, y tick style={color={rgb,1:red,0.0;green,0.0;blue,0.0}, opacity={1.0}}, y tick label style={color={rgb,1:red,0.0;green,0.0;blue,0.0}, opacity={1.0}, rotate={0}}, ylabel style={at={(ticklabel cs:0.5)}, anchor=near ticklabel, font={{\fontsize{11 pt}{14.3 pt}\selectfont}}, color={rgb,1:red,0.0;green,0.0;blue,0.0}, draw opacity={1.0}, rotate={0.0}}, ymajorgrids={false}, ymin={-0.00429}, ymax={0.14729}, ytick={{0.0,0.03,0.06,0.09,0.12}}, yticklabels={{$0.00$,$0.03$,$0.06$,$0.09$,$0.12$}}, ytick align={inside}, yticklabel style={font={{\fontsize{8 pt}{10.4 pt}\selectfont}}, color={rgb,1:red,0.0;green,0.0;blue,0.0}, draw opacity={1.0}, rotate={0.0}}, y grid style={color={rgb,1:red,0.0;green,0.0;blue,0.0}, draw opacity={0.1}, line width={0.5}, solid}, axis y line*={left}, y axis line style={color={rgb,1:red,0.0;green,0.0;blue,0.0}, draw opacity={1.0}, line width={1}, solid}]
    \addplot[color={rgb,1:red,0.0;green,0.0;blue,0.0}, name path={626292e9-a4ce-45f2-a4ad-3231acf53aea}, draw opacity={1.0}, line width={1}, solid]
        table[row sep={\\}]
        {
            \\
            100.0  0.1234  \\
            110.0  0.0779  \\
            120.0  0.050199999999999995  \\
            130.0  0.026699999999999998  \\
            140.0  0.0127  \\
            150.0  0.0075  \\
            160.0  0.0014000000000000002  \\
            170.0  0.0063  \\
            180.0  0.0044  \\
            190.0  0.0027999999999999995  \\
            200.0  0.0005  \\
            210.0  0.0017000000000000001  \\
            220.0  0.0016  \\
        }
        ;
    \addplot[color={rgb,1:red,0.0;green,0.0;blue,0.0}, name path={18db80f8-005f-4d6d-8819-8baeeca39f1d}, draw opacity={1.0}, line width={1}, solid, mark={-}, mark size={3.0 pt}, mark repeat={1}, mark options={color={rgb,1:red,0.0;green,0.0;blue,0.0}, draw opacity={1.0}, fill={rgb,1:red,0.0;green,0.0;blue,0.0}, fill opacity={1.0}, line width={0.75}, rotate={0}, solid}]
        table[row sep={\\}]
        {
            \\
            100.0  0.1047  \\
            100.0  0.14300000000000002  \\
        }
        ;
    \addplot[color={rgb,1:red,0.0;green,0.0;blue,0.0}, name path={18db80f8-005f-4d6d-8819-8baeeca39f1d}, draw opacity={1.0}, line width={1}, solid, mark={-}, mark size={3.0 pt}, mark repeat={1}, mark options={color={rgb,1:red,0.0;green,0.0;blue,0.0}, draw opacity={1.0}, fill={rgb,1:red,0.0;green,0.0;blue,0.0}, fill opacity={1.0}, line width={0.75}, rotate={0}, solid}]
        table[row sep={\\}]
        {
            \\
            110.0  0.061599999999999995  \\
            110.0  0.095  \\
        }
        ;
    \addplot[color={rgb,1:red,0.0;green,0.0;blue,0.0}, name path={18db80f8-005f-4d6d-8819-8baeeca39f1d}, draw opacity={1.0}, line width={1}, solid, mark={-}, mark size={3.0 pt}, mark repeat={1}, mark options={color={rgb,1:red,0.0;green,0.0;blue,0.0}, draw opacity={1.0}, fill={rgb,1:red,0.0;green,0.0;blue,0.0}, fill opacity={1.0}, line width={0.75}, rotate={0}, solid}]
        table[row sep={\\}]
        {
            \\
            120.0  0.03769750000000002  \\
            120.0  0.0635  \\
        }
        ;
    \addplot[color={rgb,1:red,0.0;green,0.0;blue,0.0}, name path={18db80f8-005f-4d6d-8819-8baeeca39f1d}, draw opacity={1.0}, line width={1}, solid, mark={-}, mark size={3.0 pt}, mark repeat={1}, mark options={color={rgb,1:red,0.0;green,0.0;blue,0.0}, draw opacity={1.0}, fill={rgb,1:red,0.0;green,0.0;blue,0.0}, fill opacity={1.0}, line width={0.75}, rotate={0}, solid}]
        table[row sep={\\}]
        {
            \\
            130.0  0.0194  \\
            130.0  0.0344  \\
        }
        ;
    \addplot[color={rgb,1:red,0.0;green,0.0;blue,0.0}, name path={18db80f8-005f-4d6d-8819-8baeeca39f1d}, draw opacity={1.0}, line width={1}, solid, mark={-}, mark size={3.0 pt}, mark repeat={1}, mark options={color={rgb,1:red,0.0;green,0.0;blue,0.0}, draw opacity={1.0}, fill={rgb,1:red,0.0;green,0.0;blue,0.0}, fill opacity={1.0}, line width={0.75}, rotate={0}, solid}]
        table[row sep={\\}]
        {
            \\
            140.0  0.0071  \\
            140.0  0.0191  \\
        }
        ;
    \addplot[color={rgb,1:red,0.0;green,0.0;blue,0.0}, name path={18db80f8-005f-4d6d-8819-8baeeca39f1d}, draw opacity={1.0}, line width={1}, solid, mark={-}, mark size={3.0 pt}, mark repeat={1}, mark options={color={rgb,1:red,0.0;green,0.0;blue,0.0}, draw opacity={1.0}, fill={rgb,1:red,0.0;green,0.0;blue,0.0}, fill opacity={1.0}, line width={0.75}, rotate={0}, solid}]
        table[row sep={\\}]
        {
            \\
            150.0  0.0038000000000000004  \\
            150.0  0.0119  \\
        }
        ;
    \addplot[color={rgb,1:red,0.0;green,0.0;blue,0.0}, name path={18db80f8-005f-4d6d-8819-8baeeca39f1d}, draw opacity={1.0}, line width={1}, solid, mark={-}, mark size={3.0 pt}, mark repeat={1}, mark options={color={rgb,1:red,0.0;green,0.0;blue,0.0}, draw opacity={1.0}, fill={rgb,1:red,0.0;green,0.0;blue,0.0}, fill opacity={1.0}, line width={0.75}, rotate={0}, solid}]
        table[row sep={\\}]
        {
            \\
            160.0  0.0002999999999999999  \\
            160.0  0.0028000000000000004  \\
        }
        ;
    \addplot[color={rgb,1:red,0.0;green,0.0;blue,0.0}, name path={18db80f8-005f-4d6d-8819-8baeeca39f1d}, draw opacity={1.0}, line width={1}, solid, mark={-}, mark size={3.0 pt}, mark repeat={1}, mark options={color={rgb,1:red,0.0;green,0.0;blue,0.0}, draw opacity={1.0}, fill={rgb,1:red,0.0;green,0.0;blue,0.0}, fill opacity={1.0}, line width={0.75}, rotate={0}, solid}]
        table[row sep={\\}]
        {
            \\
            170.0  0.0026  \\
            170.0  0.0108  \\
        }
        ;
    \addplot[color={rgb,1:red,0.0;green,0.0;blue,0.0}, name path={18db80f8-005f-4d6d-8819-8baeeca39f1d}, draw opacity={1.0}, line width={1}, solid, mark={-}, mark size={3.0 pt}, mark repeat={1}, mark options={color={rgb,1:red,0.0;green,0.0;blue,0.0}, draw opacity={1.0}, fill={rgb,1:red,0.0;green,0.0;blue,0.0}, fill opacity={1.0}, line width={0.75}, rotate={0}, solid}]
        table[row sep={\\}]
        {
            \\
            180.0  0.0012000000000000001  \\
            180.0  0.0083  \\
        }
        ;
    \addplot[color={rgb,1:red,0.0;green,0.0;blue,0.0}, name path={18db80f8-005f-4d6d-8819-8baeeca39f1d}, draw opacity={1.0}, line width={1}, solid, mark={-}, mark size={3.0 pt}, mark repeat={1}, mark options={color={rgb,1:red,0.0;green,0.0;blue,0.0}, draw opacity={1.0}, fill={rgb,1:red,0.0;green,0.0;blue,0.0}, fill opacity={1.0}, line width={0.75}, rotate={0}, solid}]
        table[row sep={\\}]
        {
            \\
            190.0  0.0007000000000000001  \\
            190.0  0.0055000000000000005  \\
        }
        ;
    \addplot[color={rgb,1:red,0.0;green,0.0;blue,0.0}, name path={18db80f8-005f-4d6d-8819-8baeeca39f1d}, draw opacity={1.0}, line width={1}, solid, mark={-}, mark size={3.0 pt}, mark repeat={1}, mark options={color={rgb,1:red,0.0;green,0.0;blue,0.0}, draw opacity={1.0}, fill={rgb,1:red,0.0;green,0.0;blue,0.0}, fill opacity={1.0}, line width={0.75}, rotate={0}, solid}]
        table[row sep={\\}]
        {
            \\
            200.0  0.0  \\
            200.0  0.0013  \\
        }
        ;
    \addplot[color={rgb,1:red,0.0;green,0.0;blue,0.0}, name path={18db80f8-005f-4d6d-8819-8baeeca39f1d}, draw opacity={1.0}, line width={1}, solid, mark={-}, mark size={3.0 pt}, mark repeat={1}, mark options={color={rgb,1:red,0.0;green,0.0;blue,0.0}, draw opacity={1.0}, fill={rgb,1:red,0.0;green,0.0;blue,0.0}, fill opacity={1.0}, line width={0.75}, rotate={0}, solid}]
        table[row sep={\\}]
        {
            \\
            210.0  0.0002000000000000001  \\
            210.0  0.0037  \\
        }
        ;
    \addplot[color={rgb,1:red,0.0;green,0.0;blue,0.0}, name path={18db80f8-005f-4d6d-8819-8baeeca39f1d}, draw opacity={1.0}, line width={1}, solid, mark={-}, mark size={3.0 pt}, mark repeat={1}, mark options={color={rgb,1:red,0.0;green,0.0;blue,0.0}, draw opacity={1.0}, fill={rgb,1:red,0.0;green,0.0;blue,0.0}, fill opacity={1.0}, line width={0.75}, rotate={0}, solid}]
        table[row sep={\\}]
        {
            \\
            220.0  0.0  \\
            220.0  0.0041  \\
        }
        ;
\end{axis}
\end{tikzpicture}

%% file: fig/fig_loc.tex
\begin{tikzpicture}[/tikz/background rectangle/.style={fill={rgb,1:red,1.0;green,1.0;blue,1.0}, draw opacity={1.0}}, show background rectangle]
\begin{axis}[point meta max={nan}, point meta min={nan}, legend cell align={left}, title={}, title style={at={{(0.5,1)}}, anchor={south}, font={{\fontsize{14 pt}{18.2 pt}\selectfont}}, color={rgb,1:red,0.0;green,0.0;blue,0.0}, draw opacity={1.0}, rotate={0.0}}, legend style={color={rgb,1:red,0.0;green,0.0;blue,0.0}, draw opacity={1.0}, line width={1}, solid, fill={rgb,1:red,1.0;green,1.0;blue,1.0}, fill opacity={1.0}, text opacity={1.0}, font={{\fontsize{8 pt}{10.4 pt}\selectfont}}, text={rgb,1:red,0.0;green,0.0;blue,0.0}, at={(0.02, 0.98)}, anchor={north west}}, axis background/.style={fill={rgb,1:red,1.0;green,1.0;blue,1.0}, opacity={1.0}}, anchor={north west}, xshift={1.0mm}, yshift={-1.0mm}, width={99.6mm}, height={74.2mm}, scaled x ticks={false}, xlabel={$\textrm{Localization Precision (nm)}$}, x tick style={color={rgb,1:red,0.0;green,0.0;blue,0.0}, opacity={1.0}}, x tick label style={color={rgb,1:red,0.0;green,0.0;blue,0.0}, opacity={1.0}, rotate={0}}, xlabel style={at={(ticklabel cs:0.5)}, anchor=near ticklabel, font={{\fontsize{11 pt}{14.3 pt}\selectfont}}, color={rgb,1:red,0.0;green,0.0;blue,0.0}, draw opacity={1.0}, rotate={0.0}}, xmajorgrids={false}, xmin={1.3408471440969194}, xmax={7.841669650978466}, xtick={{2.0,3.0,4.0,5.0,6.0,7.0}}, xticklabels={{$2$,$3$,$4$,$5$,$6$,$7$}}, xtick align={inside}, xticklabel style={font={{\fontsize{8 pt}{10.4 pt}\selectfont}}, color={rgb,1:red,0.0;green,0.0;blue,0.0}, draw opacity={1.0}, rotate={0.0}}, x grid style={color={rgb,1:red,0.0;green,0.0;blue,0.0}, draw opacity={0.1}, line width={0.5}, solid}, axis x line*={left}, x axis line style={color={rgb,1:red,0.0;green,0.0;blue,0.0}, draw opacity={1.0}, line width={1}, solid}, scaled y ticks={false}, ylabel={$\textrm{Error Rate}$}, y tick style={color={rgb,1:red,0.0;green,0.0;blue,0.0}, opacity={1.0}}, y tick label style={color={rgb,1:red,0.0;green,0.0;blue,0.0}, opacity={1.0}, rotate={0}}, ylabel style={at={(ticklabel cs:0.5)}, anchor=near ticklabel, font={{\fontsize{11 pt}{14.3 pt}\selectfont}}, color={rgb,1:red,0.0;green,0.0;blue,0.0}, draw opacity={1.0}, rotate={0.0}}, ymajorgrids={false}, ymin={-0.0026749999999999994}, ymax={0.09184166666666665}, ytick={{0.0,0.02,0.04,0.06,0.08}}, yticklabels={{$0.00$,$0.02$,$0.04$,$0.06$,$0.08$}}, ytick align={inside}, yticklabel style={font={{\fontsize{8 pt}{10.4 pt}\selectfont}}, color={rgb,1:red,0.0;green,0.0;blue,0.0}, draw opacity={1.0}, rotate={0.0}}, y grid style={color={rgb,1:red,0.0;green,0.0;blue,0.0}, draw opacity={0.1}, line width={0.5}, solid}, axis y line*={left}, y axis line style={color={rgb,1:red,0.0;green,0.0;blue,0.0}, draw opacity={1.0}, line width={1}, solid}]
    \addplot[color={rgb,1:red,0.0;green,0.0;blue,0.0}, name path={fe8913a5-d363-4aae-803d-a1e6283cfff7}, draw opacity={1.0}, line width={1}, dotted]
        table[row sep={\\}]
        {
            \\
            1.5279680024095244  0.0  \\
            2.2934235559715535  0.0003333333333333333  \\
            3.0586952161213254  0.006833333333333334  \\
            3.818585239952839  0.007833333333333335  \\
            4.584319136399142  0.009333333333333334  \\
            5.349058086363716  0.01966666666666667  \\
            6.130844582617209  0.0365  \\
            6.870356765240793  0.06066666666666666  \\
            7.632884361909095  0.06816666666666665  \\
        }
        ;
    \addlegendentry {$200\textrm{ rounds}$}
    \addplot[color={rgb,1:red,0.0;green,0.0;blue,0.0}, name path={17a14091-f50c-4b16-a1b5-0399b4caf0a1}, draw opacity={1.0}, line width={1}, dotted, mark={-}, mark size={3.0 pt}, mark repeat={1}, mark options={color={rgb,1:red,0.0;green,0.0;blue,0.0}, draw opacity={1.0}, fill={rgb,1:red,0.0;green,0.0;blue,0.0}, fill opacity={1.0}, line width={0.75}, rotate={0}, solid}, forget plot]
        table[row sep={\\}]
        {
            \\
            1.5279680024095244  0.0  \\
            1.5279680024095244  0.0  \\
        }
        ;
    \addplot[color={rgb,1:red,0.0;green,0.0;blue,0.0}, name path={17a14091-f50c-4b16-a1b5-0399b4caf0a1}, draw opacity={1.0}, line width={1}, dotted, mark={-}, mark size={3.0 pt}, mark repeat={1}, mark options={color={rgb,1:red,0.0;green,0.0;blue,0.0}, draw opacity={1.0}, fill={rgb,1:red,0.0;green,0.0;blue,0.0}, fill opacity={1.0}, line width={0.75}, rotate={0}, solid}, forget plot]
        table[row sep={\\}]
        {
            \\
            2.2934235559715535  0.0  \\
            2.2934235559715535  0.001  \\
        }
        ;
    \addplot[color={rgb,1:red,0.0;green,0.0;blue,0.0}, name path={17a14091-f50c-4b16-a1b5-0399b4caf0a1}, draw opacity={1.0}, line width={1}, dotted, mark={-}, mark size={3.0 pt}, mark repeat={1}, mark options={color={rgb,1:red,0.0;green,0.0;blue,0.0}, draw opacity={1.0}, fill={rgb,1:red,0.0;green,0.0;blue,0.0}, fill opacity={1.0}, line width={0.75}, rotate={0}, solid}, forget plot]
        table[row sep={\\}]
        {
            \\
            3.0586952161213254  0.0028333333333333335  \\
            3.0586952161213254  0.011500000000000003  \\
        }
        ;
    \addplot[color={rgb,1:red,0.0;green,0.0;blue,0.0}, name path={17a14091-f50c-4b16-a1b5-0399b4caf0a1}, draw opacity={1.0}, line width={1}, dotted, mark={-}, mark size={3.0 pt}, mark repeat={1}, mark options={color={rgb,1:red,0.0;green,0.0;blue,0.0}, draw opacity={1.0}, fill={rgb,1:red,0.0;green,0.0;blue,0.0}, fill opacity={1.0}, line width={0.75}, rotate={0}, solid}, forget plot]
        table[row sep={\\}]
        {
            \\
            3.818585239952839  0.003  \\
            3.818585239952839  0.013666666666666667  \\
        }
        ;
    \addplot[color={rgb,1:red,0.0;green,0.0;blue,0.0}, name path={17a14091-f50c-4b16-a1b5-0399b4caf0a1}, draw opacity={1.0}, line width={1}, dotted, mark={-}, mark size={3.0 pt}, mark repeat={1}, mark options={color={rgb,1:red,0.0;green,0.0;blue,0.0}, draw opacity={1.0}, fill={rgb,1:red,0.0;green,0.0;blue,0.0}, fill opacity={1.0}, line width={0.75}, rotate={0}, solid}, forget plot]
        table[row sep={\\}]
        {
            \\
            4.584319136399142  0.004333333333333333  \\
            4.584319136399142  0.015000000000000001  \\
        }
        ;
    \addplot[color={rgb,1:red,0.0;green,0.0;blue,0.0}, name path={17a14091-f50c-4b16-a1b5-0399b4caf0a1}, draw opacity={1.0}, line width={1}, dotted, mark={-}, mark size={3.0 pt}, mark repeat={1}, mark options={color={rgb,1:red,0.0;green,0.0;blue,0.0}, draw opacity={1.0}, fill={rgb,1:red,0.0;green,0.0;blue,0.0}, fill opacity={1.0}, line width={0.75}, rotate={0}, solid}, forget plot]
        table[row sep={\\}]
        {
            \\
            5.349058086363716  0.011333333333333334  \\
            5.349058086363716  0.028833333333333336  \\
        }
        ;
    \addplot[color={rgb,1:red,0.0;green,0.0;blue,0.0}, name path={17a14091-f50c-4b16-a1b5-0399b4caf0a1}, draw opacity={1.0}, line width={1}, dotted, mark={-}, mark size={3.0 pt}, mark repeat={1}, mark options={color={rgb,1:red,0.0;green,0.0;blue,0.0}, draw opacity={1.0}, fill={rgb,1:red,0.0;green,0.0;blue,0.0}, fill opacity={1.0}, line width={0.75}, rotate={0}, solid}, forget plot]
        table[row sep={\\}]
        {
            \\
            6.130844582617209  0.025333333333333333  \\
            6.130844582617209  0.049499999999999995  \\
        }
        ;
    \addplot[color={rgb,1:red,0.0;green,0.0;blue,0.0}, name path={17a14091-f50c-4b16-a1b5-0399b4caf0a1}, draw opacity={1.0}, line width={1}, dotted, mark={-}, mark size={3.0 pt}, mark repeat={1}, mark options={color={rgb,1:red,0.0;green,0.0;blue,0.0}, draw opacity={1.0}, fill={rgb,1:red,0.0;green,0.0;blue,0.0}, fill opacity={1.0}, line width={0.75}, rotate={0}, solid}, forget plot]
        table[row sep={\\}]
        {
            \\
            6.870356765240793  0.045499999999999985  \\
            6.870356765240793  0.07783333333333332  \\
        }
        ;
    \addplot[color={rgb,1:red,0.0;green,0.0;blue,0.0}, name path={17a14091-f50c-4b16-a1b5-0399b4caf0a1}, draw opacity={1.0}, line width={1}, dotted, mark={-}, mark size={3.0 pt}, mark repeat={1}, mark options={color={rgb,1:red,0.0;green,0.0;blue,0.0}, draw opacity={1.0}, fill={rgb,1:red,0.0;green,0.0;blue,0.0}, fill opacity={1.0}, line width={0.75}, rotate={0}, solid}, forget plot]
        table[row sep={\\}]
        {
            \\
            7.632884361909095  0.048833333333333326  \\
            7.632884361909095  0.08916666666666666  \\
        }
        ;
    \addplot[color={rgb,1:red,0.0;green,0.0;blue,0.0}, name path={bbf4ca46-d6eb-49a4-b188-d88acf41a49c}, draw opacity={1.0}, line width={1}, solid]
        table[row sep={\\}]
        {
            \\
            1.5248326867445103  0.0  \\
            2.2933727225699303  0.0  \\
            3.054770198673627  0.0  \\
            3.8199027486417534  0.0003333333333333333  \\
            4.578746752455054  0.003333333333333333  \\
            5.346461605542606  0.010833333333333334  \\
            6.111306006409955  0.005500000000000001  \\
            6.866241055100078  0.019833333333333335  \\
            7.6576841083308755  0.02066666666666667  \\
        }
        ;
    \addlegendentry {$300\textrm{ rounds}$}
    \addplot[color={rgb,1:red,0.0;green,0.0;blue,0.0}, name path={cacb8b2a-85af-4802-9ed3-cde790e4ec23}, draw opacity={1.0}, line width={1}, solid, mark={-}, mark size={3.0 pt}, mark repeat={1}, mark options={color={rgb,1:red,0.0;green,0.0;blue,0.0}, draw opacity={1.0}, fill={rgb,1:red,0.0;green,0.0;blue,0.0}, fill opacity={1.0}, line width={0.75}, rotate={0}, solid}, forget plot]
        table[row sep={\\}]
        {
            \\
            1.5248326867445103  0.0  \\
            1.5248326867445103  0.0  \\
        }
        ;
    \addplot[color={rgb,1:red,0.0;green,0.0;blue,0.0}, name path={cacb8b2a-85af-4802-9ed3-cde790e4ec23}, draw opacity={1.0}, line width={1}, solid, mark={-}, mark size={3.0 pt}, mark repeat={1}, mark options={color={rgb,1:red,0.0;green,0.0;blue,0.0}, draw opacity={1.0}, fill={rgb,1:red,0.0;green,0.0;blue,0.0}, fill opacity={1.0}, line width={0.75}, rotate={0}, solid}, forget plot]
        table[row sep={\\}]
        {
            \\
            2.2933727225699303  0.0  \\
            2.2933727225699303  0.0  \\
        }
        ;
    \addplot[color={rgb,1:red,0.0;green,0.0;blue,0.0}, name path={cacb8b2a-85af-4802-9ed3-cde790e4ec23}, draw opacity={1.0}, line width={1}, solid, mark={-}, mark size={3.0 pt}, mark repeat={1}, mark options={color={rgb,1:red,0.0;green,0.0;blue,0.0}, draw opacity={1.0}, fill={rgb,1:red,0.0;green,0.0;blue,0.0}, fill opacity={1.0}, line width={0.75}, rotate={0}, solid}, forget plot]
        table[row sep={\\}]
        {
            \\
            3.054770198673627  0.0  \\
            3.054770198673627  0.0  \\
        }
        ;
    \addplot[color={rgb,1:red,0.0;green,0.0;blue,0.0}, name path={cacb8b2a-85af-4802-9ed3-cde790e4ec23}, draw opacity={1.0}, line width={1}, solid, mark={-}, mark size={3.0 pt}, mark repeat={1}, mark options={color={rgb,1:red,0.0;green,0.0;blue,0.0}, draw opacity={1.0}, fill={rgb,1:red,0.0;green,0.0;blue,0.0}, fill opacity={1.0}, line width={0.75}, rotate={0}, solid}, forget plot]
        table[row sep={\\}]
        {
            \\
            3.8199027486417534  0.0  \\
            3.8199027486417534  0.001  \\
        }
        ;
    \addplot[color={rgb,1:red,0.0;green,0.0;blue,0.0}, name path={cacb8b2a-85af-4802-9ed3-cde790e4ec23}, draw opacity={1.0}, line width={1}, solid, mark={-}, mark size={3.0 pt}, mark repeat={1}, mark options={color={rgb,1:red,0.0;green,0.0;blue,0.0}, draw opacity={1.0}, fill={rgb,1:red,0.0;green,0.0;blue,0.0}, fill opacity={1.0}, line width={0.75}, rotate={0}, solid}, forget plot]
        table[row sep={\\}]
        {
            \\
            4.578746752455054  0.0005  \\
            4.578746752455054  0.007333333333333332  \\
        }
        ;
    \addplot[color={rgb,1:red,0.0;green,0.0;blue,0.0}, name path={cacb8b2a-85af-4802-9ed3-cde790e4ec23}, draw opacity={1.0}, line width={1}, solid, mark={-}, mark size={3.0 pt}, mark repeat={1}, mark options={color={rgb,1:red,0.0;green,0.0;blue,0.0}, draw opacity={1.0}, fill={rgb,1:red,0.0;green,0.0;blue,0.0}, fill opacity={1.0}, line width={0.75}, rotate={0}, solid}, forget plot]
        table[row sep={\\}]
        {
            \\
            5.346461605542606  0.005833333333333333  \\
            5.346461605542606  0.01666666666666667  \\
        }
        ;
    \addplot[color={rgb,1:red,0.0;green,0.0;blue,0.0}, name path={cacb8b2a-85af-4802-9ed3-cde790e4ec23}, draw opacity={1.0}, line width={1}, solid, mark={-}, mark size={3.0 pt}, mark repeat={1}, mark options={color={rgb,1:red,0.0;green,0.0;blue,0.0}, draw opacity={1.0}, fill={rgb,1:red,0.0;green,0.0;blue,0.0}, fill opacity={1.0}, line width={0.75}, rotate={0}, solid}, forget plot]
        table[row sep={\\}]
        {
            \\
            6.111306006409955  0.0025  \\
            6.111306006409955  0.009000000000000001  \\
        }
        ;
    \addplot[color={rgb,1:red,0.0;green,0.0;blue,0.0}, name path={cacb8b2a-85af-4802-9ed3-cde790e4ec23}, draw opacity={1.0}, line width={1}, solid, mark={-}, mark size={3.0 pt}, mark repeat={1}, mark options={color={rgb,1:red,0.0;green,0.0;blue,0.0}, draw opacity={1.0}, fill={rgb,1:red,0.0;green,0.0;blue,0.0}, fill opacity={1.0}, line width={0.75}, rotate={0}, solid}, forget plot]
        table[row sep={\\}]
        {
            \\
            6.866241055100078  0.0125  \\
            6.866241055100078  0.028004166666666615  \\
        }
        ;
    \addplot[color={rgb,1:red,0.0;green,0.0;blue,0.0}, name path={cacb8b2a-85af-4802-9ed3-cde790e4ec23}, draw opacity={1.0}, line width={1}, solid, mark={-}, mark size={3.0 pt}, mark repeat={1}, mark options={color={rgb,1:red,0.0;green,0.0;blue,0.0}, draw opacity={1.0}, fill={rgb,1:red,0.0;green,0.0;blue,0.0}, fill opacity={1.0}, line width={0.75}, rotate={0}, solid}, forget plot]
        table[row sep={\\}]
        {
            \\
            7.6576841083308755  0.013500000000000002  \\
            7.6576841083308755  0.02866666666666667  \\
        }
        ;
\end{axis}
\end{tikzpicture}